\documentstyle[twocolumn]{mn}

\def\beq{\begin{equation}}
\def\eeq{\end{equation}}
\def\bey{\begin{eqnarray}}
\def\eey{\end{eqnarray}}

\def\MJ{M_{\rm Jup}}

\def\br{{\bf r}}
\def\Ploc{{P_{\rm loc}}}
\def\Pobs{P_{\rm obs}}

\def\atan{{\rm atan}}

\def\spose#1{\hbox to 0pt{#1\hss}}
\def\lta{\mathrel{\spose{\lower 3pt\hbox{$\sim$}}
    \raise 2.0pt\hbox{$<$}}}
\def\gta{\mathrel{\spose{\lower 3pt\hbox{$\sim$}}
    \raise 2.0pt\hbox{$>$}}}

\input epsf

\title[Biases in Cometary Catalogues and Planet X]
	{Biases in Cometary Catalogues and Planet X}

\author[]
        {}

\author[J. Horner \& N.W. Evans]
       {J. Horner \& N.W. Evans\\
        Theoretical Physics, 1 Keble Rd, Oxford, OX1 3NP}

\date{Accepted ........      Received .......;      in original form .......}
\pagerange{\pageref{firstpage}--\pageref{lastpage}}
\pubyear{2001}

\begin{document}
\maketitle
\label{firstpage}
\maketitle
\label{firstpage}

\begin{abstract}
Two sets of investigators -- Murray (1999) and Matese, Whitman \&
Whitmire (1999) -- have recently claimed evidence for an undiscovered
Solar System planet from possible great circle alignments in the
aphelia directions of the long period comets. However, comet
discoveries are bedevilled by selection effects. These include
anomalies caused by the excess of observers in the northern as against
the southern hemisphere, seasonal and diurnal biases, directional
effects which make it harder to discover comets in certain regions of
the sky, as well as sociological biases. A simple mathematical model
is developed to illustrate the geometrical selection effects
controlling comet discoveries.  

The stream proposed by Murray is shown on an equal area Hammer-Aitoff
projection. The addition of newer data weakens the case for the
alignment. There is also evidence that the subsample in the stream is
affected by seasonal and north-south biases.  The stream proposed by
Matese et al. is most obvious in the sample of dynamically new comets,
and especially in those whose orbits are best known. The most recent
data continues to maintain the overpopulation in the great circle.
This pattern in the data occurs with a probability of only $\sim 1.5 \times
10^{-3}$ by chance.  None of the known biases are able to provide such
an alignment.  Numerical integrations are used to demonstrate that a
planet by itself can reduce the perihelia of comets in its orbital
plane to sufficiently small values so that they could be discovered
from the Earth. To maintain the observed flux of comets in the stream
requires a parent population of $\sim 3 \times 10^{9}$ objects on
orbits close to the planet's orbital plane.

There is a need for a sample of long period comets that is free from
unknown or hard-to-model selection effects. Such will be provided by
the European Space Agency satellite GAIA, which will discover $\sim
1000$ long period comets during its 5 year mission.  This may finally
bring to fruition the long tradition of looking for the effects of
perturbers in cometary catalogues.
\end{abstract}


\begin{figure*}
\epsfysize=9.cm \centerline{\epsfbox{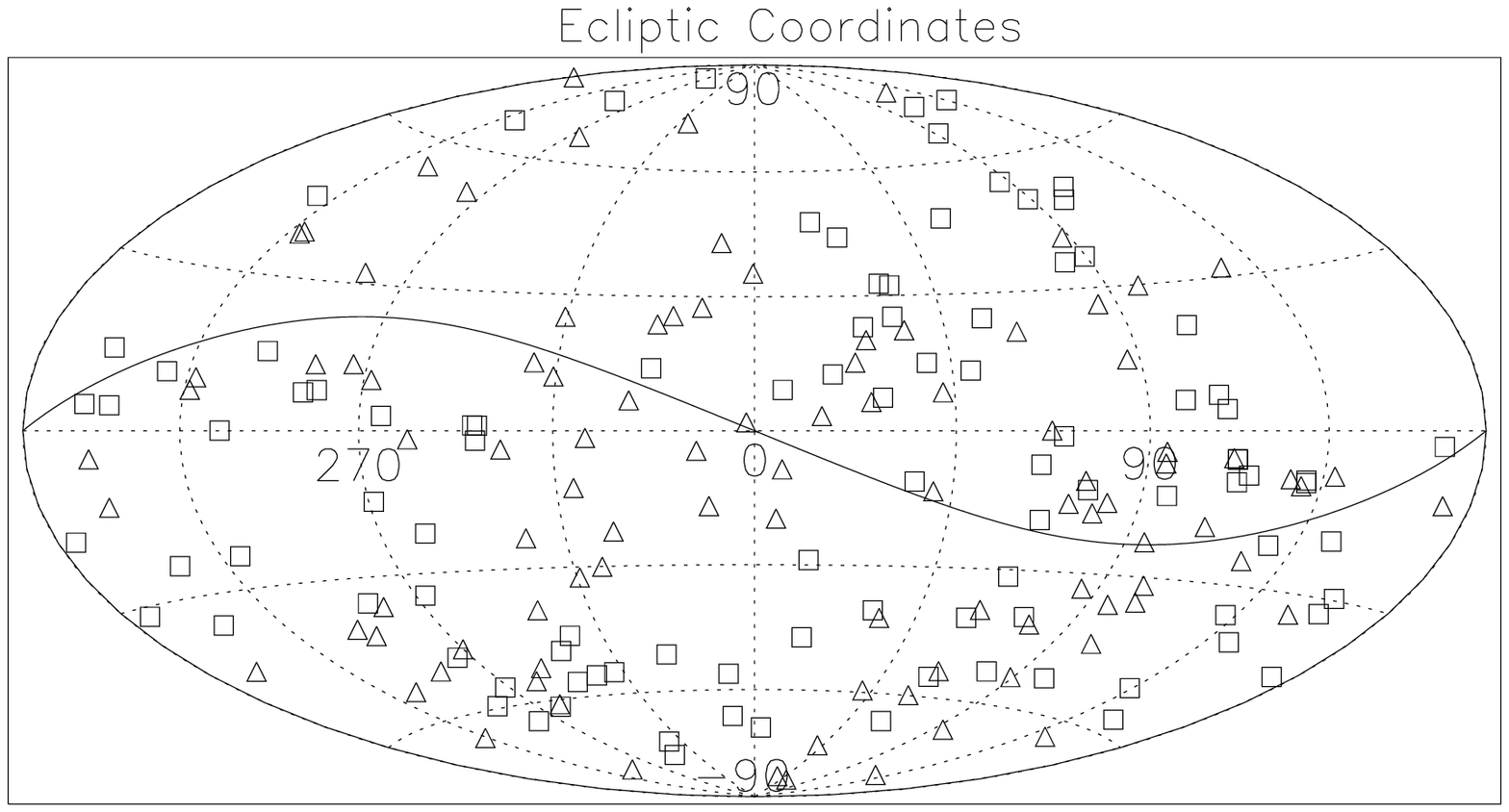}}
\vskip-1.0truecm
\epsfysize=9.cm \centerline{\epsfbox{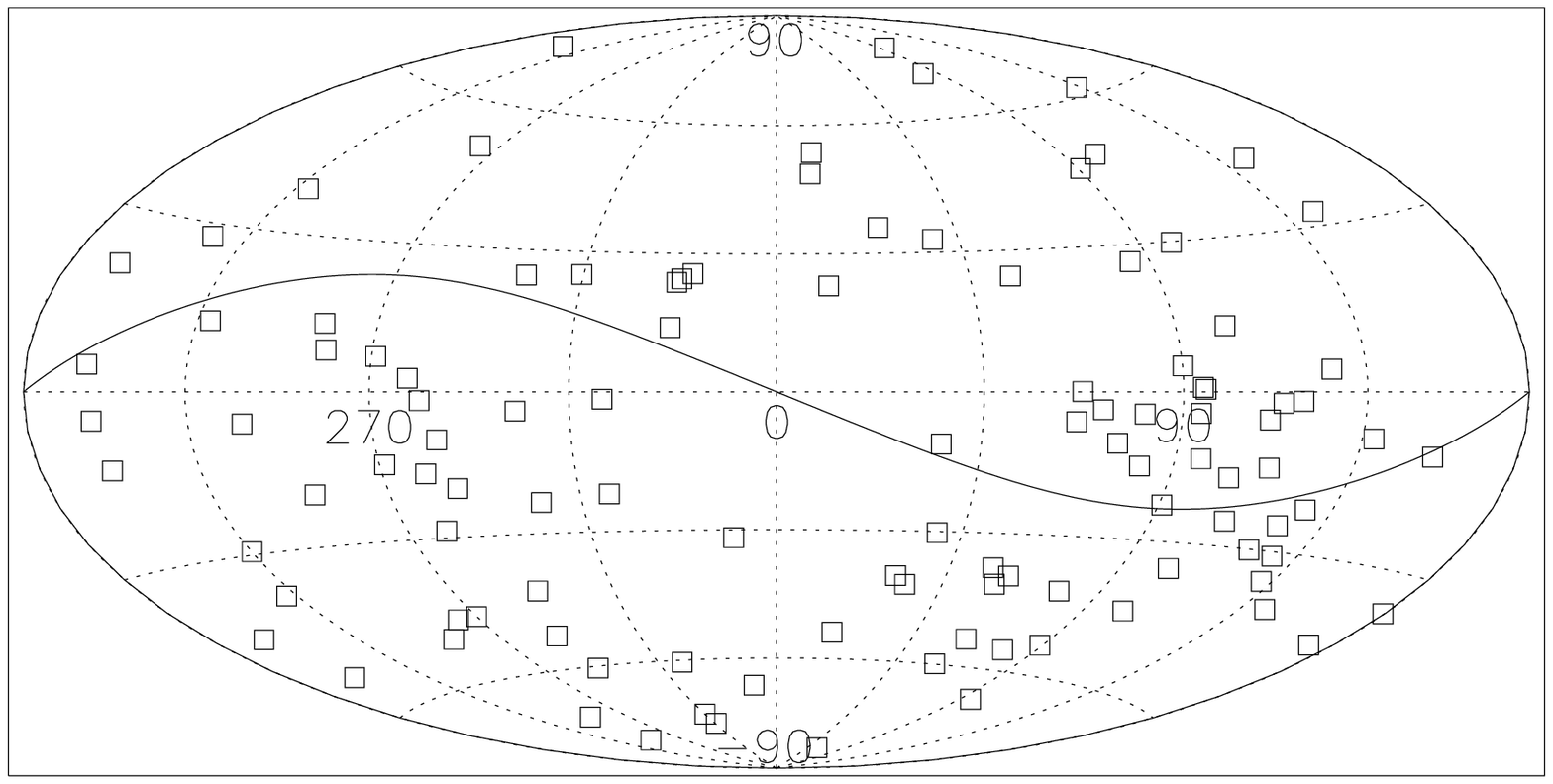}}
\caption{Upper panel: Plot of the aphelia of all 1A (squares) and 1B
(triangles) comets in Marsden \& Williams (1999) catalogue in
heliocentric ecliptic longitude and latitude. Meridians of latitude
are shown at $30^\circ$ intervals, parallels of longitude at
$45^\circ$ intervals. The centre of the projection corresponds to the
zero of longitude and is the first point of Aries. By convention,
increasing longitude runs to the right, which is the mirror image of
the sky as viewed by an observer on Earth.  The solid line shows the
projection of the Earth's equator onto the celestial sphere. Lower
panel: Plot of the aphelia of all (1A, 1B, 2A and 2B) comets
discovered prior to 1940. The deficiency of comets with northern
aphelia, especially at high latitude, is immediately apparent.}
\label{fig:allcomets}
\end{figure*}

\begin{table*}
\begin{center}
\begin{tabular}{|l|c|c|c|c|c|c|} \hline
Discovery & Northern   & Southern & Problematic & Problematic &
 Problematic & Total \\
Hemisphere & Perihelia & Perihelia & Probably North & Probably South & \null & \\ \hline
North  & $71$ & $17$  & $10$ & $12$ & $13$ & $123$ \\ \hline 
South  & $6$  & $17$  & $1$  & $2$  & $2$ & $28$ \\ \hline
\end{tabular}
\end{center}
\caption{Table showing discovery hemisphere versus hemisphere of
perihelion passage for the 151 comets in Marsden \& Williams (1999)
which also appear in Vsekhsvyatskii (1964). The table demonstrates
that comets passing perihelion in the celestial Northern hemisphere
are preferentially discovered by observers in the terrestrial northern
hemisphere, and vice versa. (Problematic comets are those difficult to
classify either because they are closer than $15^\circ$ to the Sun at
perihelion or because they pass perihelion on or near the celestial
equator. If a tentative hemisphere of perihelion can be assigned to a
problematic comet, we have tried to do so.)}
\label{table:jontisstats}
\end{table*}

\section{Introduction}

Since the discovery of Pluto, the idea that there may be an
undiscovered tenth planet (Planet X) lying still further from the Sun
has often been put forward.  Attempts to predict the existence of
undiscovered planets from studies of the aphelia of periodic comets
have been made many times (e.g., Forbes 1880; Kritzinger 1957; Opik
1971; Guliev 1992).  Most of these authors argued for a close planet
at $\sim 50$ astronomical units (au). There is also work motivated by
periodicities in the cratering and fossil records of $\approx 30$
Myr. Here, the idea is that the advance or the regression of the
perihelion or aphelion of Planet X through the Kuiper Belt triggers
periodic comet showers (e.g., Whitmire \& Matese 1985; Matese \&
Whitmire 1986).  Assuming Planet X is responsible for the
periodicities, then it has a semimajor axis of typically $\sim 100$ au
and a mass of a few Earth masses. A different hypothesis, motivated by
the same periodicities, is a distant unseen companion to the Sun
(Nemesis) on an eccentric orbit with semimajor axis $\sim 90000$ au
and period $\sim 27$ Myr (Whitmire \& Jackson 1984; Davis et
al. 1984). The orbit has an assumed perihelion of $\sim 20000$ au,
close enough to make incursions into the inner Oort Cloud and trigger
comet showers (see Tremaine (1986), Vandervoort \& Sather (1993), for
reviews of the status of these theories).

The idea has most recently surfaced in this {\it Journal} in Murray
(1999), who claimed first that the aphelia of long period comets
between $30000$ and $50000$ show an excess and second that they are
aligned along a great circle inclined at $\sim 30^\circ$ to the
ecliptic. Murray interpreted this as evidence of the imprint of a body
orbiting the Sun at $\sim 32000$ au with a period of $5.8 \times 10^6$
years. Planet X is presumably massive (to produce a detectable family
of comets) and its orbit is possibly retrograde, suggesting a recent
capture by the Sun.  A similar analysis was carried out by Matese,
Whitman \& Whitmire (1999), who looked at the same dataset of aphelia
of long period comets. Matese et al. (1999) identified an excess of
aphelia on a different great circle, this time with galactic longitude
$\sim 135^\circ$, and argued that this must be produced by a bound
perturber in the Oort Cloud.  They suggested that there was a planet
or brown dwarf with mass $\sim 3 \MJ$ at $\sim 25000$ au on an orbit
roughly normal to the Galactic plane. The distances suggested by
Murray and Matese et al. for Planet X are in good agreement, but the
comets that belong to the great circle streams are quite different in
the two papers, as is the orientation of the orbit.

Even so, there is a some supporting theoretical evidence for Planet
X. First, studies of the ejection and growth of planetesimals in the
early Solar System do suggest the existence of a number of small
planets beyond Pluto (e.g., Stern 1991, Kenyon \& Luu 1999). Second,
there are unexplained residuals in the motions of Uranus and Neptune,
although it is difficult to evaluate the trustworthiness of the
historical data (e.g., Anderson \& Standish 1986, Hogg, Quinlan \&
Tremaine 1991).

In this paper, we analyse the evidence put forward by Murray (1999)
and Matese et al. (1999). We begin with a thorough examination of the
biases that afflict catalogues of long period comets in Section 2 and
develop a mathematical model of the biases in Section 3 to illustrate
our main points. Section 4 returns to the great circle alignments
found by Murray (1999) and Matese et al. (1999) and assesses whether
they are unexpected or not. Simulations are used to illustrate the
possible signal of a massive perturber. In the concluding Section 5,
we draw attention to the fact that the GAIA satellite will provide an
all-sky catalogue of comets brighter than 20th magnitude and largely
free of directional biases. We argue that this may finally bring to
fruition the long tradition of looking for the effects of planets on
the orbits of comets.

\section{Observational Biases}

Murray used the 1994 version, and Matese et al. the 1996 version, of
the catalogue of long period comets by Marsden \& Williams. Our data
are extracted from the 1999 version. It is vital that the
observational biases in this dataset be analysed before any great
circle alignments can be accepted as real. In the past, a number of
people have suggested possible effects (see e.g., Everhart 1967,
Hughes 1983, Kres\'ak 1983, R\"ust 1984, Neslu{\v s}an 1996), though a
comprehensive treatment seems lacking.

\subsection{North-South Biases}

Marsden \& Williams (1999) divide the long period comets into four
classes, namely 1A, 1B, 2A and 2B, in order of diminishing certainty
of the orbital elements.  The upper panel of
Figure~\ref{fig:allcomets} shows the aphelia of all 1A and 1B
comets. This is an equal area Hammer-Aitoff projection using
heliocentric ecliptic longitude and latitude coordinates.  The bold
line is the projection of the Earth's equator onto the celestial
sphere. It is evident from the figure that many more cometary aphelia
lie to the south rather than the north of the projection of the
Earth's equator.  Table~\ref{table:jontisstats} shows the raw
statistics on perihelion passage as seen from Earth together with the
hemisphere of discovery using comets from Marsden \& Williams and data
from Vsekhsvyatskii (1964). The table demonstrates that objects
discovered in the north generally have perihelia in the north, and
vice versa.  Of course, there is a known excess of observers in the
northern as against the southern hemisphere.  So, this bias leads to a
greater number of comets with aphelia in the southern rather than the
northern hemisphere.  Note that comets with perihelia within $\sim
30^\circ$ of the projected terrestrial equator may be discovered from
either hemisphere. Hence, this bias is most pronounced in the sample
of comets with high latitude aphelia.  The lower panel of
Figure~\ref{fig:allcomets} shows all comets (1A, 1B, 2A and 2B) in
Marsden \& Williams (1999) catalogue discovered prior to 1940. The
bias in favour of southern aphelia is very pronounced, especially at
high latitudes. This is expected because the excess of northern
hemisphere observers relative to the southern hemisphere was greatest
in the past.

Matese, Whitman \& Whitmire (1998) contend that the north-south bias
is minimized for comets whose orbits are well enough known. They argue
that these comets are primarily bright.  However, it could also be
argued that, in order for a comet be studied enough for its orbit to
be well known, it must be observed over a reasonable period of time,
and that the excess of northern hemisphere observers again means that
this is more likely to happen for comets with northern rather than
southern perihelia.

\begin{figure}
\epsfxsize=9.5cm \centerline{\epsfbox{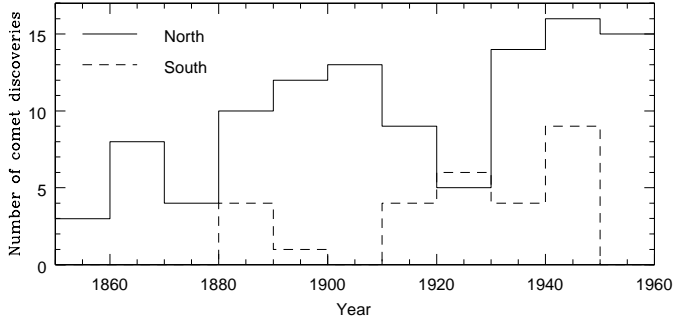}}
\caption{Histogram of the number of comets discovered from the
northern (solid lines) and southern (dotted lines) hemispheres in each
decade prior to 1956. The comets are taken from Marsden \& Williams (1999)
catalogue, but the date of discovery and location of the discoverer are
taken from Vsekhsvyatskii's (1964) book.}
\label{fig:histo}
\end{figure}

Figure~\ref{fig:histo} shows a histogram of the number of comets in
Marsden \& Williams (1999) discovered per decade from the northern and
southern hemisphere between 1810 and 1956, using the information in
Vsekhsvyatskii's (1964) book. This clearly shows the substantial bias
towards northern hemisphere observers, which is still the case as late
as the 1950s (the records in Vsekhsvyatskii's book end in 1956). It
even seems that the effects of great upheavals like the Boer War, the
First World War and the Great Depression can be discerned in the
diagram.  However, statistics are low and so the passing interests of
individual observers like Barnard, Perrine and Giacobini may also be a
contributory factor.  At first sight, it is surprising that there is
not a dip during the Second World War. Looking at all single
apparition long period comets in Vsekhsvyatskii (1964) gives us a
larger sample than in the Marsden \& Williams catalogue. We see that
the number of comets discovered in five-year intervals from 1930 to
1950 by observers based in America rises from 3 to 4 to 5 to 7. For
Southern Hemisphere observers, it changes from 3 to 2 to 6 to 13,
while for observers based in Europe, it changes from 7 to 2 to 4 to
5. This suggests that the rise in discoveries from the United States
and the Southern Hemisphere masks the deleterious effects of the
Second World War on European discoveries.  Similarly, plotting the
cumulative total of numbered asteroids against time shows a steady,
almost exponential increase, but there is a prominent plateau
coinciding with the duration of the Second World War and ending in the
1960s.  All this suggests that the cometary (and asteroidal)
catalogues are not just telling us about astronomy!

\begin{figure}
\epsfysize=5.cm \centerline{\epsfbox{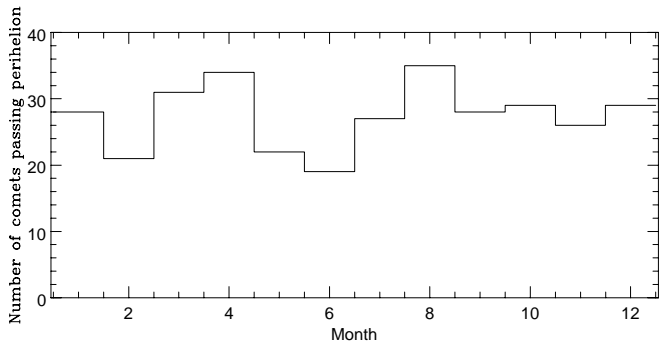}}
\caption{Histogram of the number of comets against month of perihelion
passage. This uses all comets (1A, 1B, 2A and 2B) in Marsden \&
Williams (1999) catalogue.}
\label{fig:months}
\end{figure}
\begin{figure}
\epsfysize=5.cm \centerline{\epsfbox{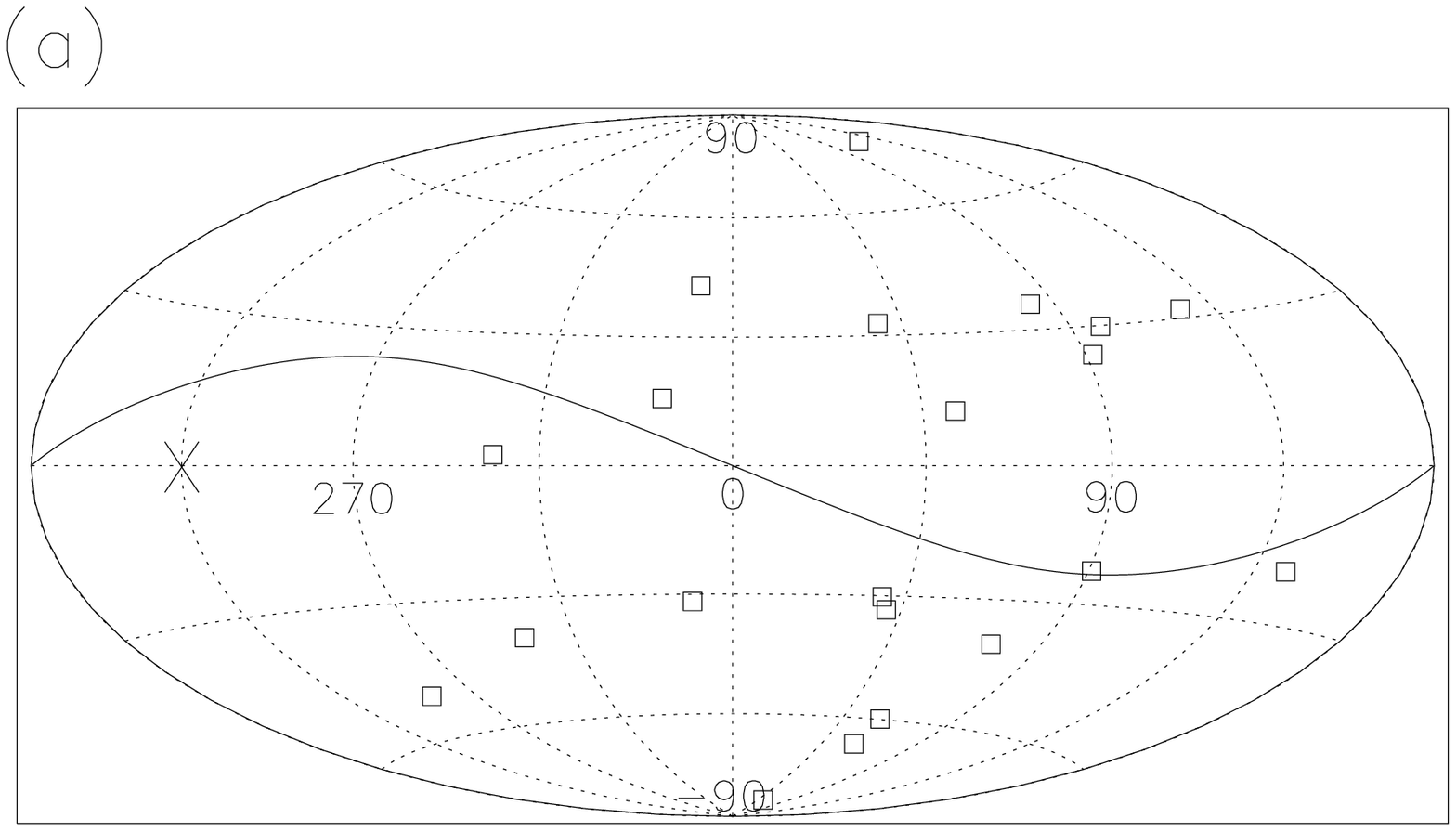}}
\epsfysize=5.cm \centerline{\epsfbox{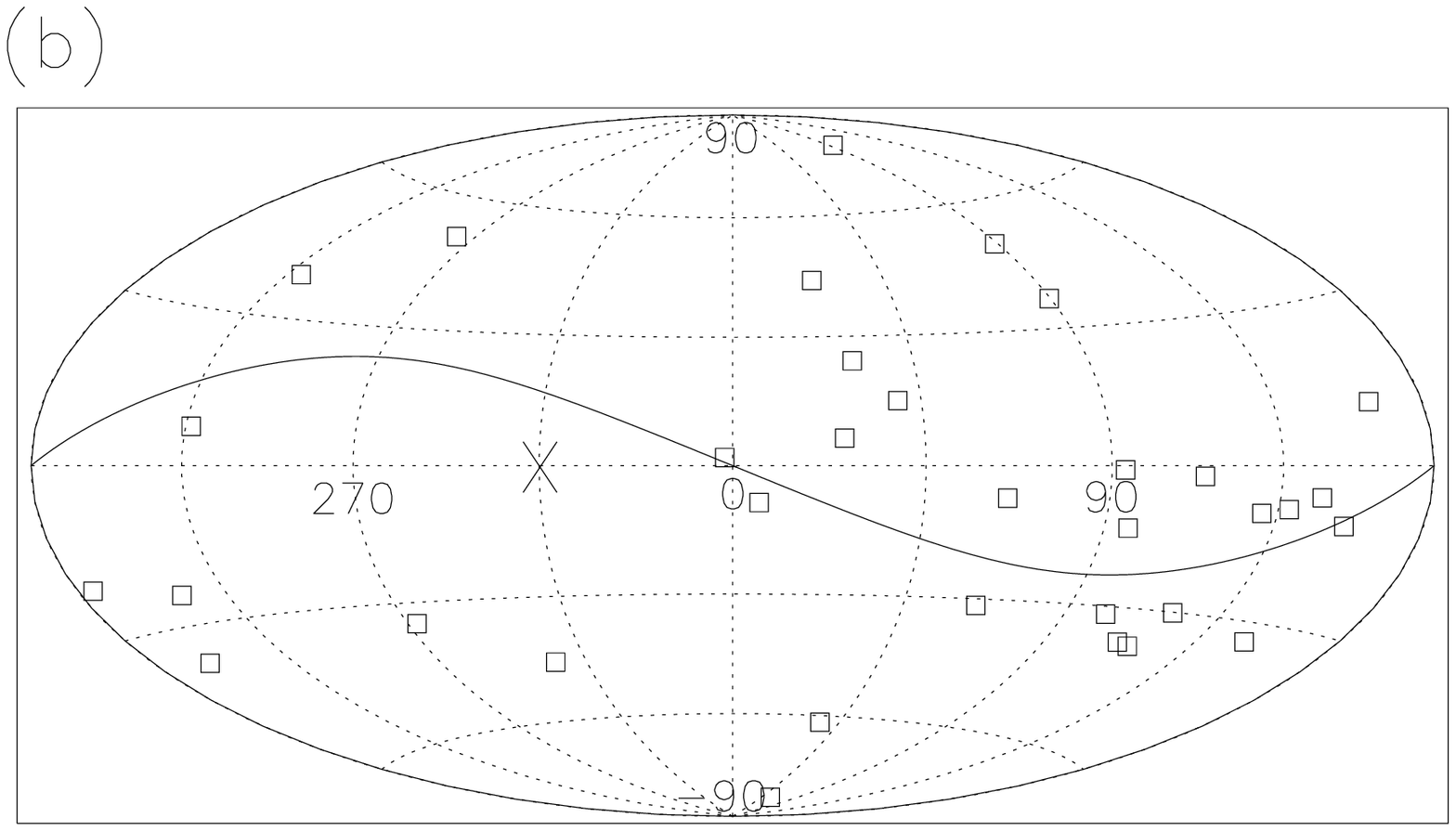}}
\epsfysize=5.cm \centerline{\epsfbox{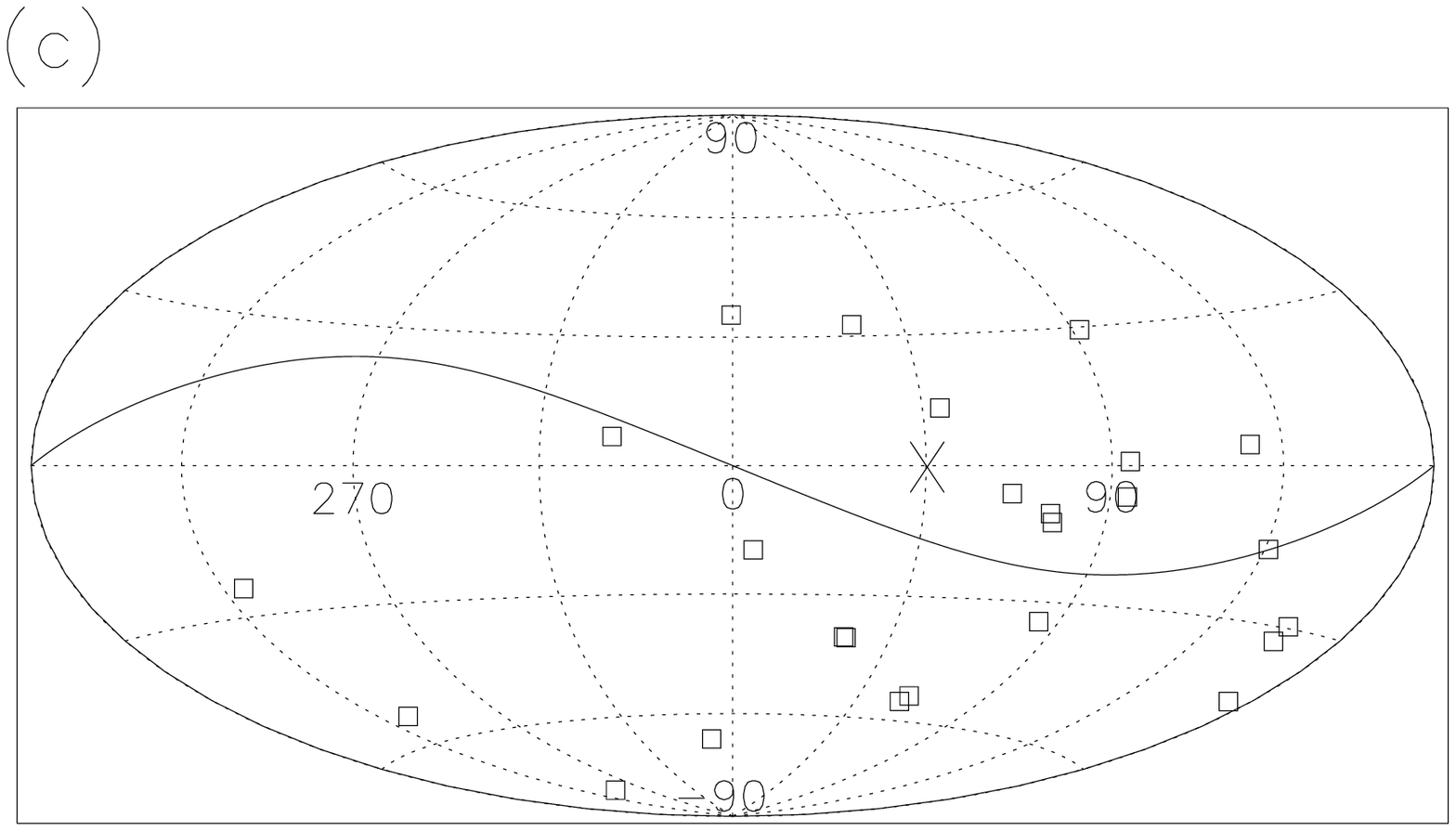}}
\epsfysize=5.cm \centerline{\epsfbox{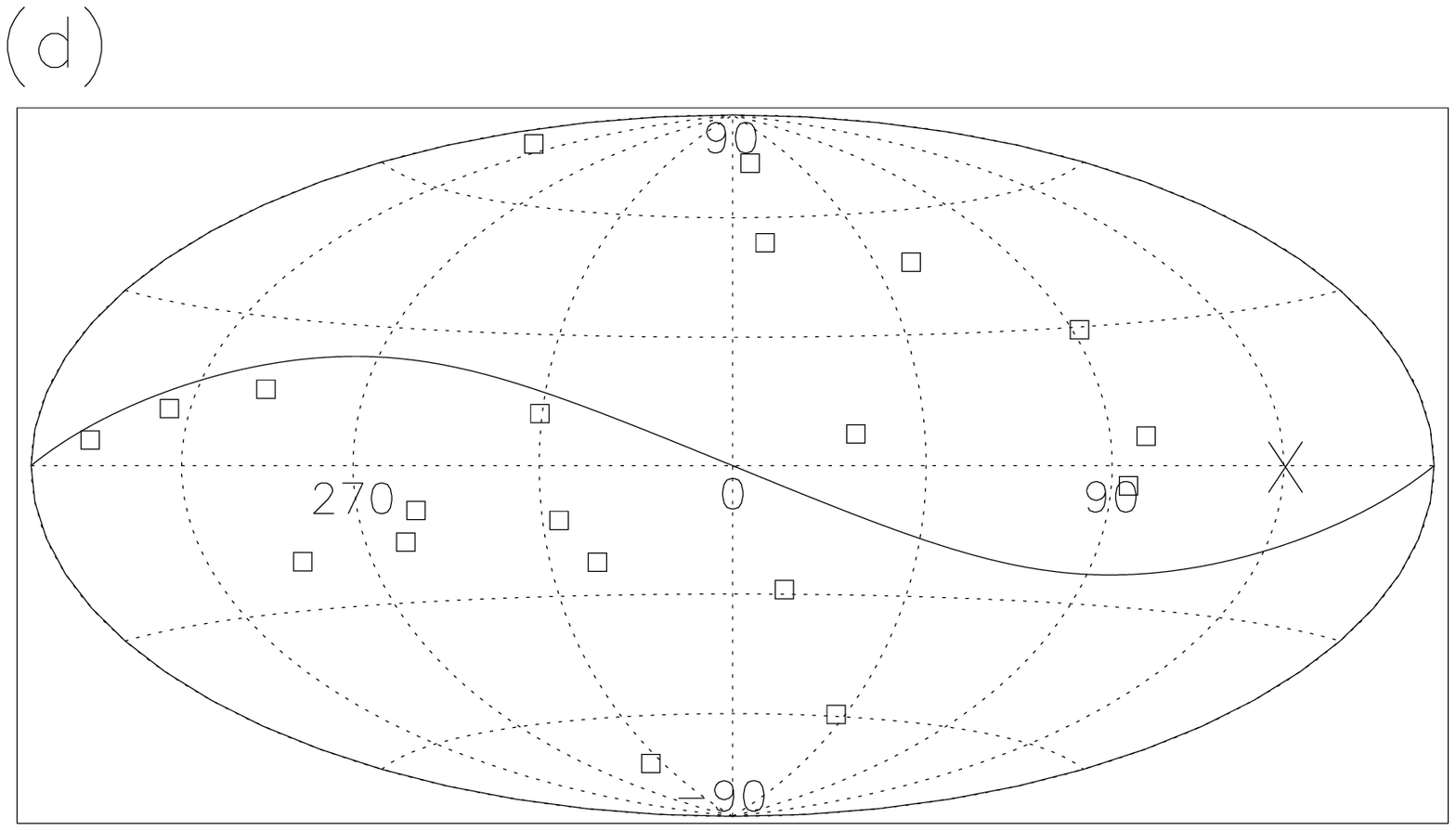}}
\caption{Plots of the aphelia of all comets (1A, 1B, 2A and 2B) in
Marsden \& Williams (1999) with perihelion passage in (a) May, (b)
August, (c) November and (d) February.  Meridians of ecliptic latitude
are shown at $30^\circ$ intervals, parallels of ecliptic longitude at
$45^\circ$ intervals. The centre of the projection is the zero of
longitude and increasing longitude runs to the right. The solid line
shows the projection of the Earth's equator onto the celestial sphere.
Notice (1) the north-south bias is most pronounced in northern winter
months and (2) there is a zone of avoidance $\approx 180^\circ$ away
from the position of the Sun (marked with a cross) as viewed from the
Earth in panels (a), (b) and (d).}
\label{fig:seasons}
\end{figure}

\subsection{Diurnal and Seasonal Biases}

There are a number of effects introduced into the catalogue of comets
due to diurnal and seasonal bias. The time of day influences the area
of the sky in which comets can be discovered. In general, a comet
cannot be discovered until it is properly dark (after astronomical
twilight), unless the comet is unusually bright or passes very close
to the Sun. Similarly, if the comet is near opposition, it is further
from the Sun than at quadrature, and is therefore harder to see. This
suggests that the distribution of comet discoveries has a minimum
towards the Sun and a secondary minimum towards midnight. All other
things being equal, we expect there to be two equal maxima, one in the
evening, and one in the morning, displaced towards midnight from the
point of quadrature due to the effects of twilight.  The seasons also
impart a bias to the data, since the length of night varies, and hence
there is more time to observe in winter, and therefore more objects
will be discovered in winter than in summer.  This can be seen to an
extent in the raw numbers from the Marsden \& Williams catalogue,
which are plotted in Figure~\ref{fig:months}.  However, the overall
effect of this is probably reduced by the fact that the weather is
generally worse in winter than summer at the temperate northern
latitudes in which cometary observers predominate, which leads to a
lowering in the number of observations made in winter. The most
obvious feature in Figure~\ref{fig:months} is a pronounced dip in
mid-summer, along with peaks in April and August. Discoveries between
August and April are roughly constant (bearing in mind that February
is 10 \% shorter than other months). The effects of the weather seem
to counterbalance the effect of longer nights in these months.

Figure~\ref{fig:seasons} shows plots of aphelia of comets which passed
through perihelion in the months of May, August, November and February
respectively.  Ideally, these plots should show the comets discovered
in each month. Unfortunately, the discovery months are scattered
through the literature for post-1956 comets. So, we assume that the
discovery took place close to perihelion.  All the comets in Marsden
\& Williams (1999) catalogue are used. This seems reasonable because
the direction of the perihelion in heliocentric coordinates should be
fairly well-known, even if the detailed orbit is subject to
uncertainties.  The north-south bias is most pronounced in the
northern winter months (see Figures~\ref{fig:seasons} (c) and
(d)). This is the coupled effect of the longer nights of the northern
hemisphere winter, together with the excess of northern observers. By
contrast, in Figures~\ref{fig:seasons} (a) and (b), which correspond
to the northern hemisphere summer months, there are very nearly equal
numbers of comets with perihelia in each terrestrial hemisphere. The
north-south bias of observers is almost cancelled by the longer nights
of the southern hemisphere winter.

\begin{figure}
\epsfysize=5.cm \centerline{\epsfbox{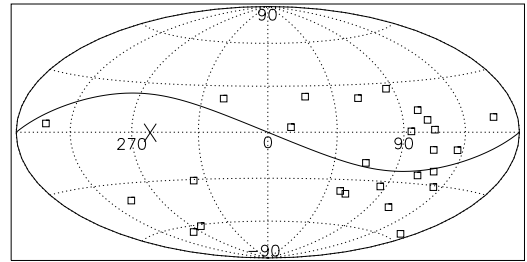}}
\caption{Plot of the aphelia of all comets (1A, 1B, 2A and 2B) in
Marsden \& Williams (1999) with perihelion passage in July.  The solid
line shows the projection of the Earth's equator onto the celestial
sphere.  Notice (1) there are no aphelia above an ecliptic latitude of
$30^\circ$ North and (2) the zone of avoidance is evident at an
ecliptic longitude of $\approx 285^\circ$ (marked with a cross).}
\label{fig:july}
\end{figure}

The panels in Figure~\ref{fig:seasons} also show a zone of avoidance
(marked with a cross) at longitudes $\approx 180^\circ$ away from the
position of the Sun as viewed from the Earth. Comets cannot be
discovered too close to the Sun. This means that comets with perihelia
close to the Sun are unlikely to be discovered near perihelion.
Holetschek (1891; see also Hughes 1983) pointed out a similar effect
-- namely, that comets with perihelion the far side of the Sun are
less likely to be discovered both because of the proximity of the Sun
and because they are further away and hence fainter.  In May, the zone
of avoidance is centered at an ecliptic longitude of $\approx
225^\circ$ and advances by $90^\circ$ as we move forward by three
months.  The zone is obvious in the May, August and February plots,
although not at all in the November plot.  Figure~\ref{fig:july} shows
the aphelia of comets which passed through perihelion in the month of
July. It is included because the zone of avoidance is very clearly
seen opposite the Sun at ecliptic longitudes of $\approx
285^\circ$. Even more strikingly, the region of sky above $30^\circ$
north of the ecliptic is completely devoid of all cometary
aphelia. This is strong evidence that even in summer, the smaller
numbers of southern observers are affecting the results.  Another
effect that is visible in the July and August plots is a tendency for
comets to cluster about the apparent position of the Sun. These are
comets with perihelia outside the Earth's orbit discovered near
opposition and possibly near perihelion. A northern hemisphere
observer at this time of year has a comparatively small area of sky to
search, due to the short summer nights. This biases cometary
discoveries to the area around opposition, which explains the apparent
clustering.

There is another diurnal effect that has been suggested in the
literature.  As is well-known, many more meteoroids are observed in
the morning as compared to the evening (e.g., McKinley 1961). We
encounter many more meteoroids on the forward side of the Earth, and
those encountered are also brighter as they have a higher relative
velocity.  By contrast, in the evening the meteoroids have to catch up
the motion of the Earth and so are scarcer and fainter. By analogy,
Kres\'ak (1983) suggested that such a diurnal asymmetry may be present
in cometary discoveries. This, however, is not clear, as comets are
not discovered through encounter with the Earth and so relative
velocities do not appear to be important. Kres\'ak's bias may have
been important in the days of naked eye discoveries, as comets that
have passed through perihelion are often brighter. This may have
biased statistics towards the discoveries of retrograde comets in the
morning. 

\subsection{Directional Biases}

There is a bias against areas of the sky with high stellar densities
or dense nebulosity. The crowded nature and high surface brightness of
such areas makes it harder to discover a comet against the
background. Hence, we might expect a dearth of cometary discoveries,
for example, in the plane of the Milky Way. However, there may be
other subtle effects operating, since some areas of the sky are of much
greater interest to observers than others. If an area of sky is
receiving twice the attention of a neighbouring area, we might assume
that the better observed area will yield a greater number of comet
discoveries.  An interesting example of a recent comet discovered in a
region with a dense stellar background is Comet Hale-Bopp (C/1995
O1). This was found independently by Hale and Bopp, who were observing
the globular cluster M70 in the direction of Sagittarius (see {\it IAU
Circular 6187}). The comet just happened to lie in the same field of
view.

Another directional bias comes from groups of comets that fragmented
at a perihelion passage prior to discovery. The aphelia of such
objects will form a tight cluster. The most obvious example is the
Kreutz family of sun-grazing comets (see e.g., Marsden 1989). There
are 3 Kreutz family sun-grazers in the catalogue of osculating
elements of Marsden \& Williams (namely, C/1882 R1-B, C/1963 R1,
C/1965 S1-A). These can be seen in the upper panel of
Figure~\ref{fig:allcomets} as the three superposed symbols at an
ecliptic longitude of $\approx 130$ degrees.  

Finally, directional biases may also come from showers of comets
produced by weak stellar perturbations of the Oort Cloud.  For
example, Biermann, Huebbner \& L\"ust (1983) pointed out that there is
a surplus of cometary aphelia in the region bounded by $180^\circ \lta
\ell \lta 240^\circ$ and $0^\circ \lta b \lta 30^\circ$, where
($\ell,b$) are galactic longitude and latitude. They attributed this
to this to an encounter with a roughly solar mass star 3 Myrs ago.
Irrespective of whether the details of the explanation are correct,
the shower does seem to be a real dynamical effect, as it is more
prominent in the more tightly bound comets (e.g., Matese et al. 1999).
Further weak showers may remain undetected in the dataset.

\subsection{Orbital Biases}

Of course, comets do not all have an equal chance of discovery, as
some orbits offer more favourable opportunities.  It is well known
that most comet searchers concentrate their efforts close to the plane
of the ecliptic, and so high inclination comets are under-represented
in catalogues. The bias towards low inclinations in long period comet
catalogues is still further enhanced by contamination of the sample
with misclassified short period comets.  Comets with very small
perihelion distances spend less time in the inner solar system than
those at large perihelion distances and cover a much smaller arc. They
are hence more likely to be missed, unless they are bright enough to
be seen when very close to the Sun. Similarly, objects with large
perihelion distance will be observable on a large arc. Comets with
perihelia lying noticeably beyond the Earth's orbit will be most
favourably placed for discovery at opposition rather than at
perihelion passage.  Comets with perihelia the far side of the Sun are
less likely to be discovered both because of the closeness of the Sun
and because they are further away and fainter (the Holetschek effect
mentioned in Section 2.2).

Comets with both high inclination and small perihelia offer particular
difficulties. For example, if the the perihelion is in the north, then
the comet spends most of its observable time in the southern
hemisphere and may well be too close to the Sun at perihelion to be
observed from the north. Table~\ref{table:jontisstats} demonstrates
that comets tend to be discovered in the same hemisphere as the
perihelion passage, but this is not true for for comets with small
perihelia ($\lta 0.2 $ au).

Finally, objects in outburst are unpredictable and may be discovered
at large heliocentric distances. Unusually bright and large comets,
such as Hale-Bopp, are likely to be discovered a long time before
perihelion passage. Comets are known to be much more active when new
and fade on subsequent apparitions, presumably because the most
volatile components are substantially lost at the first perihelion
passage leading to a dimming of the intrinsic brightness thereafter.
So, first time entrants from the Oort Cloud over the past 200 years
(the time over which records have been maintained) are more likely to
have been discovered than comets which were returning for a second or
third time during the same period.

\subsection{Sociological Biases}

The heart of the problem is the uneven coverage of the sky at
different epochs by different generations of observers.  We have
already suggested in Section 2.1 that the effects of historic
upheavals like the First World War and the Great Depression may be
able to be discerned in cometary catalogues. There may be additional
effects that are sociological in origin.  For example, especially in
observations from casual observers, there may be an additional time of
day bias. Many casual observers prefer to observe in the evening
before going to bed, rather than in the morning after rising. Due to
the difference in weather in winter and summer, casual observers in
the northern hemisphere would perhaps rather observe in the warm
summer months, which may slew the seasonal observations towards the
months after the summer solstice. This is in accord with
Figure~\ref{fig:months}, which shows that August is the best month for
cometary discoveries.

\subsection{Biases with Survey Programmes}

Since 1999, the overwhelming majority of comets have been discovered
by search programmes, such as LINEAR
(``http://www.ll.mit.edu/LINEAR/'') and NEAT
(``http://neat.jpl.nasa.gov/'').  For example, 14 of the last 20
comets in the catalogue of Marsden \& Williams were discovered by
LINEAR and a further 2 comets were found by NEAT.  It is apparent,
given the success of these programmes, that in the future such surveys
will constitute the overwhelming bulk of discoveries. The days of
comet discoveries by amateur observers are drawing to a close.  The
surveys evade some of the biases above, but they introduce their own
selection effects.  One bias immediately obvious is that both LINEAR
and NEAT are situated in the northern hemisphere. Neither will
discover comets in the far south of the sky. This is the same
situation as was present for visual observers in the past -- a
north-south bias is introduced into the dataset due to the observers
rather than any astronomical effect.  In addition, it appears from the
skyplots of observations made by LINEAR that the area around the North
Celestial Pole is not observed and so this area will show a dearth of
discoveries.  Another bias is that the survey telescopes only operate
during 'dark time'. So, the periods around full moon will not be
covered by the telescopes. This means that comet discoveries will by
bunched away from periods of moonlight to a greater extent than in the
past.

\subsection{Anisotropies}

Finally, there have been repeated claims of anisotropies in the
aphelia directions (e.g., Yabushita 1979). The preferred direction has
typically been found to coincide closely with the solar antapex.  This
played an important role historically as it was one of the pieces of
evidence that was thought to favour Lyttleton's (1953) accretion
theory of cometary origin.  The strength and direction of any
anisotropy can be quantified by harmonic analysis. This idea is
familiar in cosmic ray physics (e.g., Linsley 1975; Evans, Ferrer \&
Sarkar 2002), although apparently has not been used before in cometary
science.  Let ($\lambda, \beta$) be heliocentric ecliptic coordinates
and
\begin{equation}
N = \sum_i h(\lambda_i,\beta_i).
\label{eq:ffone}
\end{equation}
The Fourier coefficients are
\begin{equation}
a = \frac{2}{N} \sum_i h(\lambda_i,\beta_i) 
\cos \lambda_i,\quad
b = \frac{2}{N} \sum_i h(\lambda_i, \beta_i) {\sin \lambda_i},
\label{eq:fftwo}
\end{equation}
where the index $i$ runs along all the comets in the sample and
$h(\lambda,\beta)$ is the selection function.  The amplitude $R$ and
phase $\psi$ of the first harmonic is
\begin{equation}
R = (a^2 + b^2)^{1/2}, \qquad \psi = \atan \Biggl(
{b\over a} \Biggr).
\end{equation}
The probability that an amplitude higher than that measured $R$ can
arise from an underlying isotropic distribution is given by the
formula (Linsley 1975)
\begin{equation}
P ( > R) = \exp (-k_0 ), \qquad k_0 = \frac{1}{4}R^2 N.
\end{equation}
The selection function $h(\lambda,\beta)$ is both important for
analysing any alleged anisotropies and nearly impossible to model.  It
is reasonable to argue that the cometary data accumulated over long
periods of time is reasonably independent of biases in ecliptic
longitude. A crude way to cope with the north-south bias is to split
the cometary sample into those with aphelia in the northern or
southern ecliptic hemispheres, and to analyse each sample separately.
\begin{table}
\begin{center}
\begin{tabular}{|l|c|c|c|c|} \hline
Dataset & $N$ &$R$ & $\psi$ & $P(>R)$ \\ \hline
Northern aphelia & $136$ & $0.364$  & $44.2^\circ$ & $0.011$ \\ \hline 
Southern aphelia & $194$ & $0.228$  & $-84.9^\circ$ & $0.080$ \\ \hline
All  & $330$ & $0.257$  & $68.1^\circ$  & $0.004$ \\ \hline
\end{tabular}
\end{center}
\caption{Tests for isotropy on the cometary aphelia dataset. Results
are given for the whole sample, as well as for the subsamples with
aphelia in the northern ecliptic and southern ecliptic
hemispheres. The second column gives the number of comets $N$. The
third and fourth columns give the amplitude and phase, while the fifth
gives the probability that an isotropic distribution could give an
amplitude greater than that observed $R$.}
\label{table:isotropy}
\end{table}
This is performed in Table~\ref{table:isotropy} for all comets in
Marsden \& Williams catalogue, as the the direction of the aphelion in
heliocentric coordinates should be fairly well-known, irrespective of
the other orbital uncertainties. The final column of
Table~\ref{table:isotropy} shows that the probability of an isotropic
distribution yielding such large values of the amplitude $R$ is small.
From the results of Jaschek \& Vabousquet (1994), the solar antapex
lies in the direction $\lambda = 88.3^\circ$, $\beta = -49.9^\circ$
(using equinox 2000). The longitude lies within $\approx 20^\circ$ of
the phase found by the anisotropy analysis of the whole sample.
However, the phase does not point in a consistent direction for the
three samples (comets with northern aphelia, with southern aphelia and
the entire dataset) and hence, it seems that any correlation of
anisotropy with the antapex direction is untrustworthy.

\begin{figure}
\epsfxsize=8.cm \centerline{\epsfbox{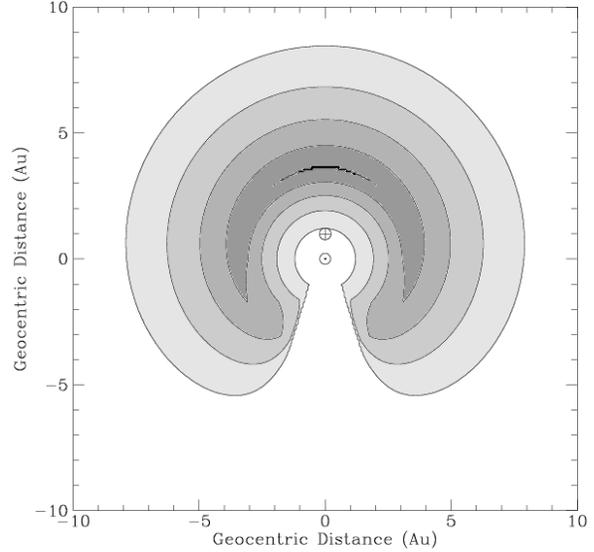}}
\caption{Contour levels of the relative probability of discovery of a
comet in the plane of the ecliptic. The positions of the Sun and the
Earth are marked. The contours are spaced at intervals of relative
probability of 0.2, with the highest level (the darkest) being
unity. The plot is drawn for a comet with absolute magnitude $H_0 = 5$
and power-law index $n=2$, together with an assumed extinction of 0.25
magnitudes per airmass.  Such a comet is five times more likely to be
discovered within the very dark crescent-shaped region than within the
lightest grey horseshoe-shaped region.}
\label{fig:prob1}
\end{figure}
\begin{figure}
\epsfxsize=8.cm \centerline{\epsfbox{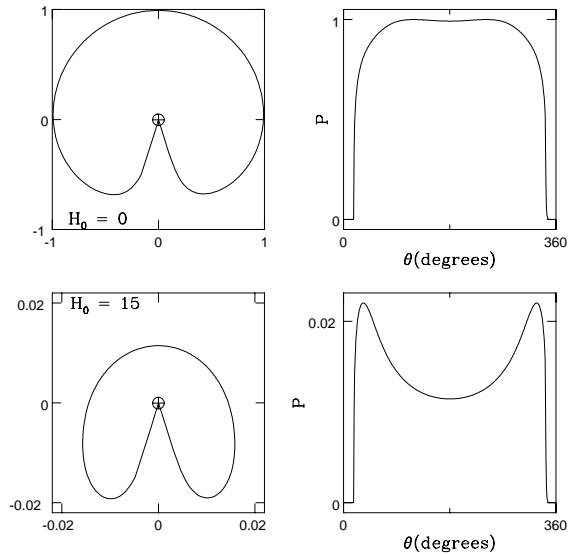}}
\caption{Plot of the relative discovery probability integrated along
the line of sight. Here, $\theta$ is measured relative to the
Earth-Sun line. In the left-hand plots, the results are shown as polar
plots with the length equal to the relative probability, whereas the
right hand plots are graphs of relative probability against
$\theta$. The upper and lower panels refer to a bright comet ($H_0 =
0$) and faint comet ($H_0 =10$) respectively. Note that for faint
comets the distribution has a minimum towards the Sun and a secondary
minimum at midnight.}
\label{fig:prob2}
\end{figure}

\section{A Model of Discovery Probability}

In this Section, we illustrate some of the biases with a simple model.
Let us denote the relative probability of discovery of a comet at any
heliocentric position $\br$ as $P(\br)$. This is equal to the relative
probability $\Ploc (\br)$ that there is a comet at the location $\br$
multiplied by the probability $\Pobs (\br)$ that a comet at $\br$
will actually be observed.

\subsection{The Algorithm}

We assume that all comets move on parabolae and that the cometary
perihelia occur uniformly in space about the Sun.  The orientation of
the orbit is completely specified if the rotation angle of the
parabola about the major axis is given. We assume that this angle is
distributed uniformly between $0^\circ$ and $360^\circ$.  We assume
that comets are discovered at random time intervals within 10 months
either side of the perihelion passage. For each cometary arc, we pick
a random time $t$ and solve for the true anomaly $v$ using (e.g.,
Danby 1992, p. 130)
\begin{equation}
{1\over 3} \tan^3 \left( {v\over2} \right) + \tan \left( {v\over 2}
\right) = {k t \over \sqrt{2q^3}},
\end{equation}
where $k$ is the Gaussian gravitational constant and $q$ is the
perihelion distance. This is used to give the heliocentric distance
$r = |\br|$ at time $t$ through
\begin{equation}
r = {q\over 1+ \cos v}.
\end{equation}
This heliocentric position corresponds to two points on the orbit,
symmetrically placed before and after perihelion.  The same
probability is allocated to each of these points.  By generating $\sim
10^6$ such trials in Monte Carlo fashion, we find that the relative
probability of a comet being at a position $r$ from the Sun is
\begin{equation}
\Ploc (\br) = {\rm min} (0.15 r^{3/2},1).
\end{equation}
The numerical pre-factor ($0.15$) varies only slightly as the window
of discovery about the perihelion point is changed. An intuitive
explanation of the power-law dependence with radius is that it is a
consequence of Kepler's Third Law.  The larger the heliocentric
distance, the more slowly a comet moves, and so the relative
probability $\Ploc$ eventually reaches a plateau.

Let $\theta$ be the angle between the comet and the Sun as seen from
the Earth. If $\theta < 18^\circ$, then the comet has set before the
end of astronomical twilight. We assume that objects so close to the
Sun are undetectable. If $\theta = 102^\circ$, then the comet is
overhead at the end of nautical twilight.  When $102^\circ < \theta
\le 180^\circ$, then the comet is directly overhead some part of the
Earth's globe during the night. Of course, the airmass is unity when
the comet is overhead.  If $18^\circ < \theta < 102^\circ$, then the
minimum extinction is calculated by working out the highest the object
can be in the sky under dark conditions. This is the zenith angle $z =
102^\circ - \theta$.  The definition of airmass $a$ is
\begin{equation}
a = {1\over \cos z}.
\end{equation}
When $z < 30^\circ$, then the airmass is computed from the slightly
more accurate formula (e.g., Kristensen 1998)
\begin{equation}
a = \sqrt{ 0.635 \cos^2 z + 2.954} - 0.79 \cos z.$$
\end{equation}
The apparent magnitude of the comet is
\begin{equation}
m = H_0 + 5 \log \Delta + 2.5n \log r + (a-1) \epsilon.
\end{equation}
Here, $r$ is the heliocentric and $\Delta$ the geocentric distance of
the comet, while $n$ is the ``power law exponent'', which determines
how rapidly the object brightens when it comes close to the Sun, and
$H_0$ is the absolute magnitude (the magnitude at 1 au from Sun and
Earth and zero phase angle). The quantity $\epsilon$ is the number of
magnitudes of extinction per airmass, which is taken as $0.25$ in our
calculations. From the apparent magnitude, we assign the probability
of observation as follows.
\begin{equation}
\Pobs = \cases{1       & if $m < 10$,\cr
                 (15-m)/5 & if $ 10 < m< 15$,\cr
                 0        & if $m > 15$.}
\end{equation}
In other words, if the comet is fainter than 15th magnitude, we assume
that it is not observed. If the comet is brighter than 10th magnitude,
the probability is unity. Between 10 and 15, the probability falls off
in a linear manner.  Of course, these choices are somewhat arbitrary,
but they are a reasonably realistic simplification of a complex
issue. Obviously, the limiting magnitudes are chosen to represent
typical conditions over the last 200 or so years of comet searching.

\begin{figure}
\epsfxsize=8.cm \centerline{\epsfbox{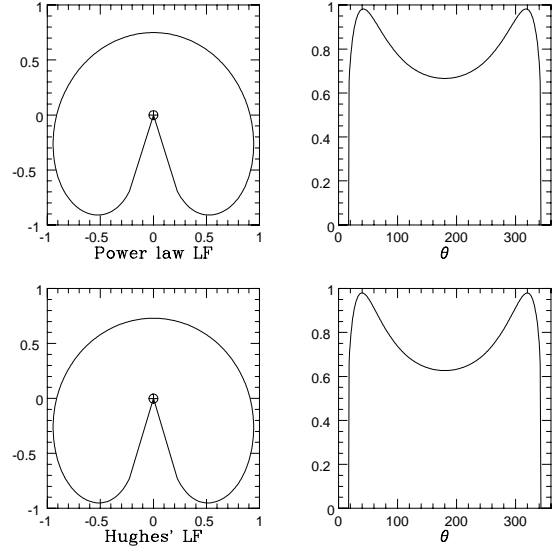}}
\caption{Plot of the relative discovery probability integrated both
along the line of sight and over the luminosity functions given in
eq~(\ref{eq:lfone}) (upper panels) and eq~(\ref{eq:lftwo}) (lower
panels). The maximum of each graph has been normalised to unity.  Note
that the secondary minimum at midnight is enhanced as compared to
Figure~\ref{fig:prob2}, as both LFs are dominated by faint objects.}
\label{fig:prob3}
\end{figure}
\begin{figure}
\epsfxsize=8.cm \centerline{\epsfbox{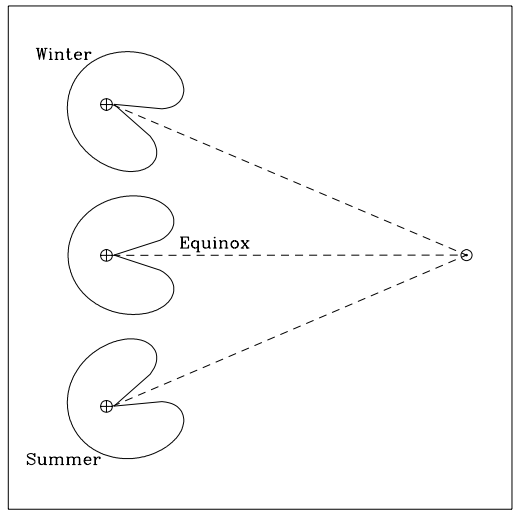}}
\caption{Plot showing how discovery probability as a function of
direction varies with the season. For a northern hemisphere observer,
the probability of discovering a comet at southerly declinations is
greatest in winter rather than summer.}
\label{fig:discseasons}
\end{figure}

\subsection{Results}

The observation probability $\Pobs$ multiplied by the location
probability $\Ploc$ gives the relative probability that a comet is
discovered at a given location. Figure~\ref{fig:prob1} shows the
probability distributions for a typical comet ($H_0 = 5$).  The figure
is plotted in the plane of the ecliptic and can be rotated about the
Sun-Earth line to give the full three-dimensional plot. The opening
angle of the two lobes is $36^\circ$, which is a consequence of the
assumption that no comets are discovered if they are so close to the
Sun that they set before the end of astronomical twilight. The
curvature of the lobes in the Sun-ward direction is controlled by the
amount of extinction -- the larger the extinction, the less likely an
object is to be discovered at high airmass and the squatter the
distribution. The overall size of the contours is set by the absolute
magnitude and power-law index of the comet.  Figure~\ref{fig:prob2}
shows the probability integrated along the line of sight as a function
of the observed angular separation of the comet from the Sun.  The
left panels show a polar and the right panels a graphical
representation of the results for a very faint ($H_0 = 15$) and very
bright ($H_0 = 0$) comet. For a comet with such characteristics, these
diagrams show the most likely angle of discovery. Note that -- as
anticipated in Section 2.2 -- the distribution has a minimum towards
the Sun and a second minimum at midnight. This effect is most obvious
in the very faint comets, as bright comets can be discovered at
greater heliocentric distances.

These figures hold good for observers on or close to the Earth's
equator (at latitudes $\lta 30^\circ$).  As we move off the equator,
two things happen. First, the angle at which the Sun sets varies from
the vertical at the equator to the horizontal at the poles. For an
object to be discovered at the poles, it would have to lie at a
declination $18^\circ$ or more to the north of the Sun. The effect of
this is to vary the opening angle of the lobes. Second, the length of
night and hence the maximum distance of the Sun beneath the horizon
also vary over the course of the year.  Again, the extreme is at the
poles, which experience $\sim 183$ days of daylight followed by $\sim
183$ days of night. This causes the overall size of the figure to
shrink or expand depending on the season.

So far, our analysis is appropriate for comets with a given value of
$H_0$ and $n$. To provide results for an ensemble, we must integrate
over the luminosity functions (LFs). We examine two LFs, the first
being a simple power-law magnitude distribution (e.g., Hughes 2001)
\begin{equation}
dN \propto 2^{H_0} dH_0, \qquad\qquad\qquad\qquad -3 < H_0< 15.
\label{eq:lfone}
\end{equation}
Here, $dN$ is the number of comets with intrinsic brightnesses in the
interval $H_0$ to $H_0 + dH_0$.  The second is the improvement
suggested by Hughes (2001), namely
\begin{equation}
dN \propto 10^{-2.607+0.359H_0} dH_0,\qquad\quad -3 < H_0 < 15.
\label{eq:lftwo}
\end{equation}
Of course, both these laws are over-simplifications and we have
extended them to magnitudes fainter than their limits of known
validity. Nonetheless, this is justifiable as we are interested in
understanding the qualitative effects of the increasing number of
objects at faint magnitudes. Comet Sarabat of 1729 is thought to have
had an absolute magnitude of -3 (e.g., Vsekhsvyatskii 1964), whilst
the faintest comets considered by Hughes have an absolute magnitude of
$\approx 15$. So we use these as our magnitude
limits. Figure~\ref{fig:prob3} shows the result of integrating over
the LFs. The results are very similar, irrespective of whether the LF
given by eq~(\ref{eq:lfone}) or eq~(\ref{eq:lftwo}) is used. Both LFs
are dominated by faint comets and so the secondary minima are very
pronounced. This result is robust against quite drastic alterations in
the LF.

Finally, Figure~\ref{fig:discseasons} illustrates the effects of the
seasons.  The position of the Earth is shown relative to the Sun at
the equinox and at the summer and winter solstices. The position of
the lobes stays constant with respect to the Earth-Sun line, which
causes them to appear to rock in declination against the background
sky to an Earth-bound observer. From the diagram, we can see that for
a northern hemisphere observer, the probability of discovering a comet
at southerly declinations is greatest in winter rather than summer.

\begin{figure}
\epsfxsize=8.cm \centerline{\epsfbox{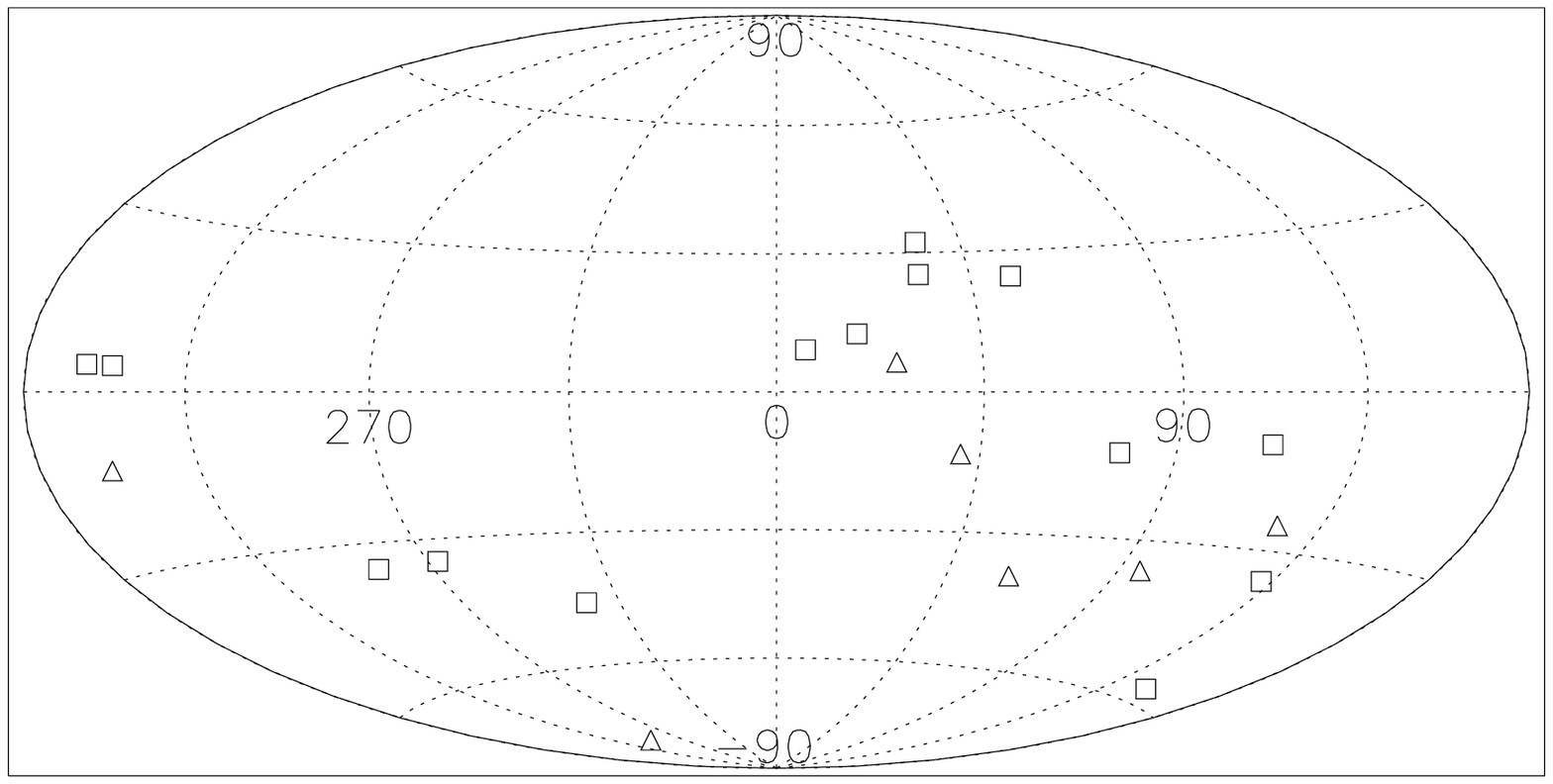}}
\epsfxsize=9.33cm \centerline{\epsfbox{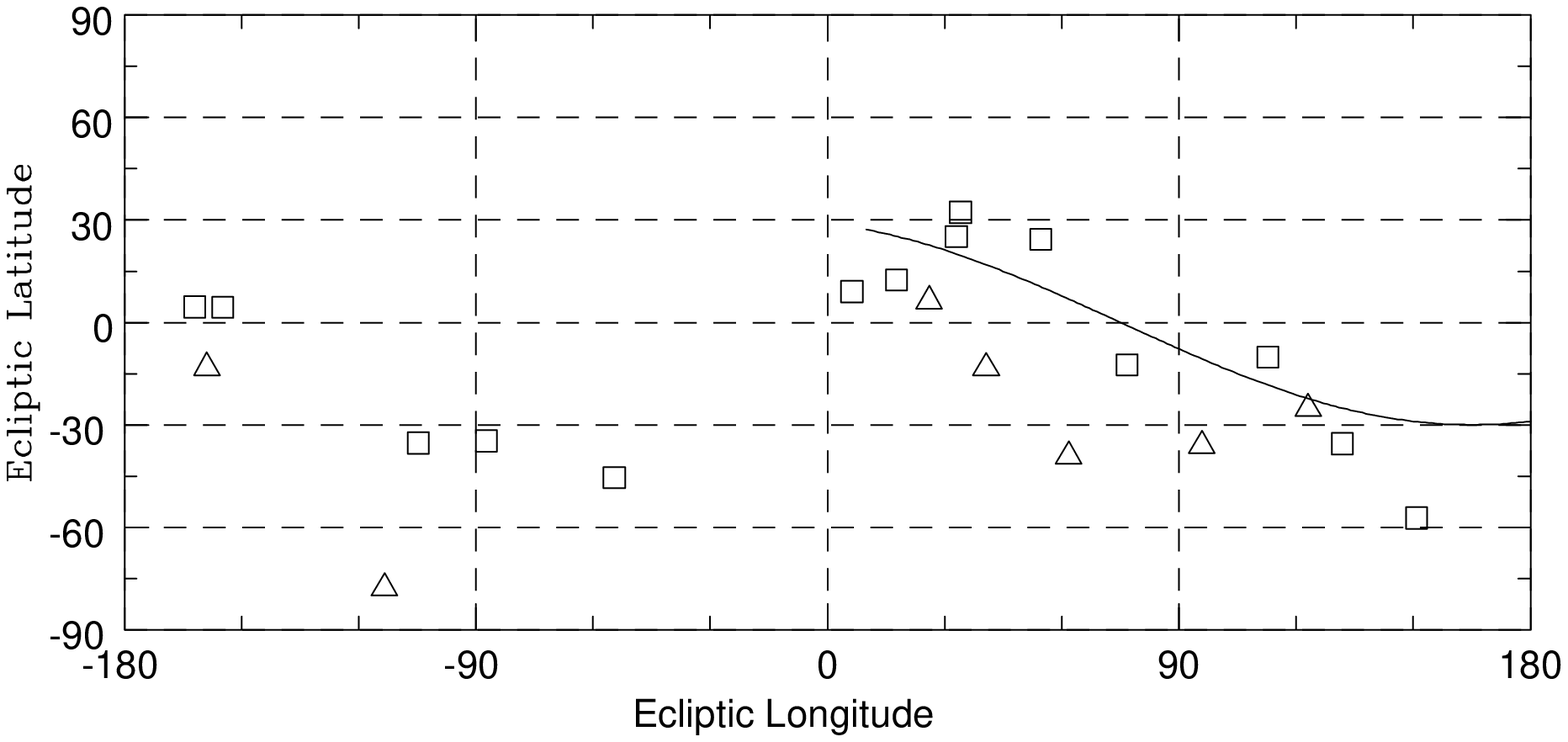}}
\caption{The upper panel shows all 1A (squares) and 1B (triangles)
comets satisfying Murray's (1999) criteria in an equal area
Hammer-Aitoff projection. The centre of the projection corresponds to
the zero of ecliptic longitude.  Meridians of ecliptic latitude are
shown at $30^\circ$ intervals, parallels of ecliptic longitude at
$45^\circ$ intervals.  The lower panel shows the data in a form
similar to that originally presented by Murray (1999), together with
the best fit sine curve to those comets identified by Murray as
objects on their first return after temporary capture by Planet
X. Note that Murray's original plot included only the 1A comets in the
1994 version of Marsden \& Williams.}
\label{fig:murray}
\end{figure}
%
\vspace{1cm} 
\begin{table*}
\begin{tabular}{|c|c|c|c|c|c|c|c|}
\hline
Name &q &$10^6/a$ &$b$ &$\ell$ &$\beta$ &$\lambda$ & Discovery \\ 
\hline\hline
C/1925 G1 & 1.109477 & +40 & 10.9 & 238.8 & -35.5 & 131.9 & Apr \\
C/1946 P1 *& 1.136113 & +44 & -19.7 & 127.2 &  32.3 &  34.1 & Aug \\
C/1989 Y1 *& 1.569172 & +49 &  -6.6 & 329.2 & -35.3 & 255.2 & Dec \\
C/1888 R1 *& 1.814916 & +48 &  59.2 & 314.9 &   4.5 & 198.0 & Sep \\
C/1932 M2 *& 2.313581 & +45 & -14.1 & 146.9 &  24.4 &  54.6 & Aug \\
C/1973 A1 & 2.511124 & +49 & -50.6 & 107.3 &   9.0 &  6.3 & Jan \\
C/1983 O1 *& 3.317899 & +48 &  54.7 & 325.3 &   4.4 & 205.2 & Jul \\ 
C/1987 F1 *& 3.624588 & +59 & -26.2 & 130.4 &  25.2 &  33.0 & Mar \\
C/1954 O2 & 3.869934 & +42 & -18.9 & 336.6 & -34.7 & 272.6 & Aug \\
C/1979 M3 & 4.686916 & +42 & 14.6 & 207.4 & -10.1 & 112.7 & Aug \\
C/1987 H1 & 5.457548 & +46 & -17.0 & 191.2 & -12.5 &  76.7 & Apr \\
C/1976 D2 *& 6.880674 & +59 & -44.2 & 121.8 &  12.4 &  17.7 & Feb \\  
\hline
C/1886 T1 & 0.663317 & +46 & -24.5 & 288.7 & -77.8 & 246.6 & Oct \\
C/1912 R1 & 0.716079 & +45 &  13.0 & 225.9 & -25.2 & 123.1 & Sep \\
C/1900 B1 *& 1.331529 & +57 &  42.1 & 305.5 & -13.2 & 201.1 & Feb \\
C/1937 C1 & 1.733791 & +62 & -38.7 & 214.7 & -39.2 &  61.7 & Feb \\
C/1990 M1 & 2.682238 & +40 & -48.4 & 172.0 & -13.2 & 40.6 & Jun \\
C/1997 J2 & 3.051136 & +40 & -0.4  & 261.4 & -57.3 & 150.8 & May \\
C/1960 M1 & 4.266927 & +40 & -12.3 & 221.2 & -36.2 & 95.9 & Jun \\
C/1999 F2 *& 4.718792 & +57 & -45.7 & 136.2 &   6.3 &  26.1 & Jun  \\
C/1993 F1 & 5.900474 & +59 & -45.5 & 330.2 & -45.5 & 305.4 & Mar\\

\hline
\hline
\end{tabular}
\caption{Data on all 1A and 1B comets satisfying Murray's (1999)
criteria. Murray considers only the 1A comets are known well enough to
qualify for membership of the great circle stream. The 12 comets in
the upper table were used in his original work. The 9 comets in the
lower table include 2 new 1A comets (C/1997 J2 and C/1993 F1) and 7 1B
comets. The table lists perihelion $q$, the reciprocal of the
semimajor axis $1/a$, galactic longitude and latitude ($\ell,b$),
ecliptic longitude and latitude ($\lambda,\beta$) and month of
discovery. All distances are in au and all angles in degrees.
(Objects with an asterisk also occur in Matese et al.'s (1999) stream
as well.) }
\label{table:murray}
\end{table*}
\begin{figure}
\epsfxsize=8.cm \centerline{\epsfbox{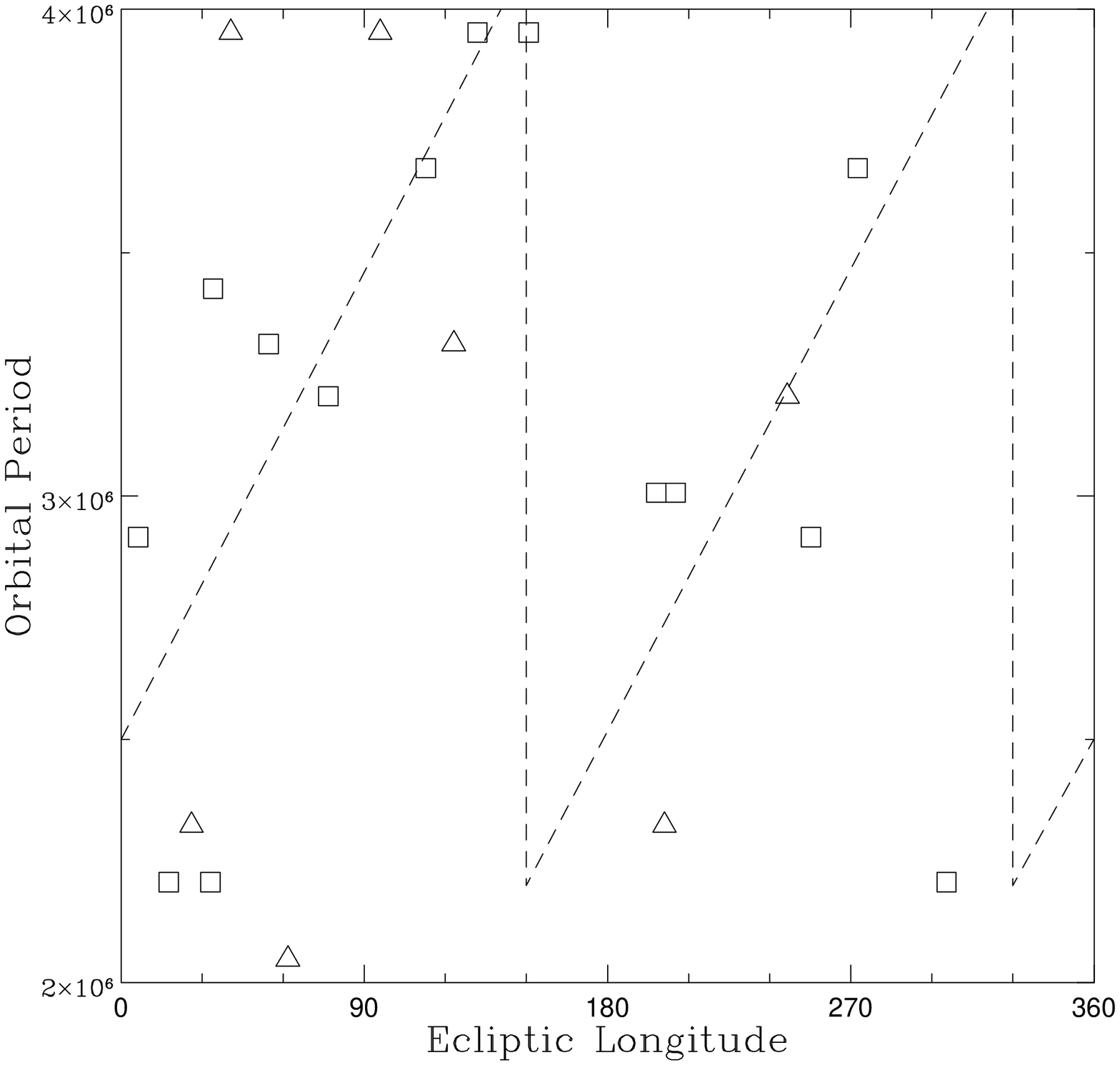}}
\caption{Plot of orbital period against ecliptic longitude for the 1A
(squares) and the 1B (triangles) comets in
Table~\ref{table:murray}. The dashed line shows the longitudinal trend
of cometary period suggested by Murray (1999). }
\label{fig:sawtooth}
\end{figure}

\section{Great Circle Alignments}

Both Murray (1999) and Matese et al. (1999) claimed the detection of
(different) great circle alignments in the aphelia of the long period
comets. In this section, we sift the evidence that they presented with
an eye as to whether biases and selection effects could have caused
the suggested alignments.

\subsection{The Planet of Murray}

First of all, Murray (1999) argued that there is an excess of aphelia
of long period comets in the range 30000 to 50000 au and that this may
be a piece of evidence for an undiscovered planet.  An immediate worry
is that for objects with aphelia $\gta 10000$ au, the eccentricity is
close to unity. Therefore, small errors in the eccentricity can cause
substantial errors in the value of $1/a$ and hence in the aphelion
distance. In the osculating elements catalogue of Marsden \& Williams
(1999), there are comets classified as 1A which are nominally
hyperbolic prior to their entry into the Solar System.  The most
extreme example is C/1996 J1-a which has a values of $1/a$ of
$-0.000511$. Even allowing for the fact that this is an unusual
object, it is clear that the typical errors in the value of $1/a$ are
sufficient to move cometary aphelia by $\sim 10000$ au at these
distances.  Another worry is that a maximum in the distribution of
aphelia seems to arise quite naturally.  Just from considerations of
spatial volume alone, the number of comets with aphelia in the range
[$Q, Q+dQ$] is expected to be an increasing function of $Q$.  In the
inner areas of the Oort Cloud ($\sim 10^3-10^4$ au), it is reasonable
to assume that very few comets will be ejected from the Cloud by the
passage of nearby stars and perturbations from the Galaxy.  The
further from the Sun, the more powerful are the effects of these
ejection mechanisms and so the lower the probability of any individual
comet surviving for the current age of the Solar System.  Hence, in
the outer parts of the Cloud, the number of comets with aphelia in the
range [$Q, Q+dQ$] is expected to be a decreasing function of $Q$.  It
is apparent from the two competing effects that a turnover or maximum
in the number of long period comets as a function of aphelion distance
is inevitable and that this maximum could well occur in the range
30000 to 50000 au.

Figure~\ref{fig:murray} shows all 1A and 1B comets fulfilling Murray's
criteria (namely, with aphelia between 30000 and 50000 au) in a
Hammer-Aitoff projection and in the Cartesian format as presented by
Murray. In fact, Murray used only 1A comets in his original paper,
arguing that these alone have sufficiently accurate orbits to be
trustworthy.  Marsden \& Williams (1999) have four classes (1A, 1B, 2A
and 2B) in order of decreasing orbital certainty and so we feel that
the 1B comets should not be entirely neglected.
Figure~\ref{fig:murray} presents the aphelia of the 1A comets as
squares and the 1Bs as triangles.  Murray's plot (pictured in the
lower panel) is not an equal area projection, but uses Cartesian
axes. This does have the advantage that the equator and any great
circle normal to the equator remain straight lines. However, it does
come at the price of serious area and direction distortion near the
poles. Any projection that is not equal area can mislead the eye in
attaching too much weight to concentrations near the equator and
rarefactions near the poles.  The comets satisfying Murray's criteria
are recorded in Table~\ref{table:murray}.  Note that our table differs
in two ways from Murray's original.  First, it incorporates 2
additional 1A comets present in the 1999, but not the 1994, version of
Marsden \& Williams catalogue. It also includes comets classified as
1Bs, on the grounds that the direction of aphelia should still be
fairly reliable, even if there are uncertainties in the other orbital
elements.  Second, one of the original comets identified by Murray for
great circle membership (C/1993 A1) is omitted because it does not
appear in Marsden \& Williams (1999).  Both the addition of the two
new 1A comets (C/1997 J2 and C/1993 F1) and the removal of C/1993 A1
weaken the case for a great circle alignment and hence a massive
object in the orbit proposed by Murray. Of the 1B comets, which have
less well-determined orbits than the 1As, 4 are positioned far from
the proposed path, 2 appear in fair agreement while 1 is in
excellent accord. Murray's stream appears clearest when the 1A comets
only are used.

There is evidence that the dataset pictured in Figure~\ref{fig:murray}
has been affected by some of the biases discussed in Section 2.  It is
apparent that most of the comets have southern aphelia rather than
northern, as a consequence of the north-south bias.  In fact, there is
only one comet with aphelion at a latitude more northerly than
$30^\circ$. It seems likely that the absence of such comets is caused
by observational selection effects.  We argued in Section 3 that
comets are usually not discovered in a band $\approx 36^\circ$ about
the Sun. The position of the Sun on the sky at the vernal equinox
(21st March) is $0^\circ$ ecliptic longitude. As the figure shows
aphelia rather than perihelia, this is inverted to $180^\circ$.  This
point moves forward by roughly $30^\circ$ per month.  Cometary
datasets are biased against this moving band.  Comets are typically
discovered 2-4 months before or after the passage of the band across
an area of sky.  The months of discovery of Murray's comets are given
in Table~\ref{table:murray}.  None of Murray's comets were discovered
in this band. So, the dataset of comets in Murray's great circle is
not entirely free from seasonal effects.

Murray's third piece of evidence is the longitudinal trend in cometary
orbital period. This is an ingenious idea which offers the possibility
of illustrating the motion of Planet X around its orbit by showing the
time at which the comets are presumed to have encountered it.
Figure~\ref{fig:sawtooth} shows our re-working of his plot, together
with Murray's original fit as a dashed line. Note that the postulated
straight lines are a simplification valid for a low inclination
planetary orbit.  The removal of C/1993 A1 from the data set detracts
somewhat from the goodness of fit.  Of the two new 1A comets, one lies
roughly in the expected position, but the other lies far away. Once
the 1B comets are added to the plot, the saw-tooth shape is even less
obvious, though it must again be noted that the orbits of these
objects are subject to more uncertainty than their 1A cousins.
Nonetheless, the technique involved in these plots is ingenious and
could be used in the future to analyse similar data for evidence
of a massive body.

Lastly, Murray claims that the bunching of all 13 aphelia of the 1A
comets into a band within $30^\circ$ of a great circle -- and so
covering only $50 \%$ of the sky -- has a probability of happening by
chance of $\sim 6 \times 10^{-4}$. This number refers to the data in
Murray's original paper. The newer data in Table~\ref{table:murray}
merely requires 14 1A comets to fall within $45^\circ$ of any great
circle. Using Monte Carlo simulations, this happens with a probability
of $\sim 0.044$, which is unusual but not astonishingly so.  Note too
that the statistical argument takes no account of selection effects.
Comets with aphelion direction close to the ecliptic are more likely to
be discovered than those close to the poles and so the dataset is more
commonplace than the statistical argument implies.

For these reasons (the alternative explanation of the bunching of
aphelia between 30000 and 50000 au; the weakening of both the sine
curve fit to the aphelia and the saw-tooth fit to the longitudinal
trend in cometary periods using newer data; the indication of
north-south and seasonal biases in the dataset), we feel that the case
for Murray's great circle alignment is no longer compelling.

\vspace{1cm} 
\begin{table}
\begin{tabular}{|l|c|c|c|c|c|c|}
\hline
Name &q &$10^6/a$ &$b$ &$\ell$ 
&$\beta$ &$\lambda$ \\ \hline
\hline
C/1958 R1 & 1.628 & +76 & -18.8 & 307.7 & -59.5 & 255.3 \\ 
C/1959 X1 & 1.253 & +69 & -78.5 & 124.3 & -19.5 &   5.0 \\ 
C/1898 L1 & 1.702 & +68 &  28.8 & 143.9 &  48.8 & 100.1 \\ 
C/1987 F1 & 3.625 & +59 & -26.2 & 130.4 &  25.2 &  33.0 \\ 
C/1976 D2 & 6.881 & +59 & -44.2 & 121.8 &  12.4 &  17.7 \\ 
C/1900 B1 & 1.332 & +57 &  42.1 & 305.5 & -13.2 & 201.1 \\ 
C/1989 Y1 & 1.569 & +49 &  -6.6 & 329.2 & -35.3 & 255.2 \\ 
C/1983 O1 & 3.318 & +48 &  54.7 & 325.3 &   4.4 & 205.2 \\ 
C/1888 R1 & 1.815 & +48 &  59.2 & 314.9 &   4.5 & 198.0 \\ 
C/1932 M2 & 2.314 & +45 & -14.1 & 146.9 &  24.4 &  54.6 \\ 
C/1946 P1 & 1.136 & +44 & -19.7 & 127.2 &  32.3 &  34.1 \\ 
C/1954 Y1 & 4.077 & +39 & -23.9 & 314.2 & -55.8 & 268.6 \\ 
C/1950 K1 & 2.572 & +37 & -43.4 & 138.5 &   7.3 &  28.7 \\ 
C/1976 U1 & 5.857 & +37 & -30.8 & 310.0 & -61.0 & 279.9 \\ 
C/1974 F1 & 3.011 & +36 &  22.1 & 140.5 &  49.9 &  88.9 \\ 
C/1948 T1 & 3.261 & +34 &  17.4 & 324.5 & -23.8 & 230.2 \\ 
C/1903 M1 & 0.330 & +33 & -34.5 & 320.7 & -52.5 & 288.4 \\ 
C/1978 A1 & 5.606 & +33 & -31.0 & 142.9 &  14.9 &  39.4 \\ 
C/1993 K1 & 4.839 & +33 &   2.6 & 131.2 &  47.2 &  56.5 \\ 
C/1989 X1 & 0.350 & +32 & -41.4 & 325.6 & -48.9 & 299.5 \\ 
C/1992 J1 & 3.007 & +28 & -44.2 & 316.8 & -55.1 & 304.9 \\ 
C/1978 H1 & 1.136 & +24 & -20.4 & 124.9 &  32.7 &  31.5 \\ 
C/1925 W1 & 1.566 & +24 & -21.6 & 323.9 & -46.7 & 269.9 \\ 
C/1944 K2 & 2.226 & +18 & -31.5 & 125.0 &  22.9 &  25.6 \\ 
C/1988 B1 & 5.031 & +13 & -48.0 & 316.1 & -54.7 & 311.6 \\ 
C/1974 V1 & 6.019 & +11 &  27.6 & 309.5 & -24.4 & 211.4 \\ 
C/1946 C1 & 1.724 & -13 & -60.7 & 312.6 & -50.9 & 322.0 \\ 
C/1983 O2 & 2.255 & -18 & -27.6 & 134.2 &  22.1 &  35.3 \\ 
C/1978 G2 & 6.282 & -23 & -33.9 & 126.9 &  20.1 &  25.9 \\ 
C/1899 E1 & 0.327 & -109 &  50.5 & 308.5 &  -4.8 & 199.1 \\ 
\hline
C/1997 BA6 & 3.436 & +29 &  28.0 & 123.0 & 66.7 &  91.5 \\
C/1999 F2 & 4.719 & +57 & -45.7 & 136.2 &   6.3 &  26.1 \\
C/1996 E1 & 1.359 & -42 & -22.9 & 304.4 & -64.1 & 260.5 \\
C/1988 A1 & 0.841 & +4881 &  -6.7 & 306.1 & -53.8 & 235.7 \\
C/1992 U1 & 2.313 & +423 & -19.3 & 321.0 & -48.3 & 265.3 \\
C/1969 O1-A& 1.719 & +555 & -39.1 & 316.0 & -56.4 & 296.2 \\
C/1983 N1 & 2.418 & +448 &  17.9 & 148.8 &  41.0 &  87.7 \\
C/1997 G2 & 3.085 & +2514 &  22.2 & 135.4 &  54.3 &  86.5 \\
C/1892 F1 & 1.971 & +846 & -14.2 & 315.6 & -50.7 & 254.6 \\ 
C/1988 L1 & 2.474 & +137 &  32.6 & 141.6 &  51.5 & 105.5 \\
C/1973 D1 & 1.382 & +1541 & -31.2 & 316.8 & -55.3 & 282.1 \\
C/1981 H1 & 2.458 & +703 & -53.7 & 123.4 &   3.2 &  15.0 \\
C/1957 P1 & 0.355 & +2001 &  -2.7 & 320.5 & -40.2 & 245.0 \\
C/1995 Q1 & 0.436 & +4458 & -39.7 & 126.6 &  15.1 &  22.9 \\
C/1975 N1 & 0.426 & +817 & -30.4 & 309.3 & -61.6 & 278.9 \\
C/1994 T1 & 1.845 & +663 &  -9.5 & 329.8 & -36.3 & 258.7 \\
C/1998 U5 & 1.236 & +10054 &  19.7 & 300.1 & -35.5 & 207.1 \\
C/1989 T1 & 1.047 & +9529 &  -5.9 & 320.7 & -41.9 & 248.5 \\
\hline
\hline
\end{tabular}
\caption{Data on comets possibly associated with Matese et al.'s
(1999) great circle stream.  The tables list perihelion $q$, the
reciprocal of the semimajor axis $1/a$, galactic longitude and
latitude ($\ell,b$), ecliptic longitude and latitude
($\lambda,\beta$). All distances are in au and all angles in degrees.
The upper table gives 30 dynamically new comets ($10^6/a < 100$) with
$120^\circ \le \ell \le 150^\circ$ or $300^\circ \le \ell \le
330^\circ$. These are the members of the great circle alignment as
identified by Matese et al. (1999). The lower table records all the
remaining 1A and 1B comets in Marsden \& Williams (1999) satisfying
the cut in galactic longitude alone. Three of these (C/1997 BA6,
C/1999 F2 and C/1996 E1) are also dynamically new comets.}
\label{table:matese}
\end{table}

\begin{figure}
\epsfxsize=8.cm \centerline{\epsfbox{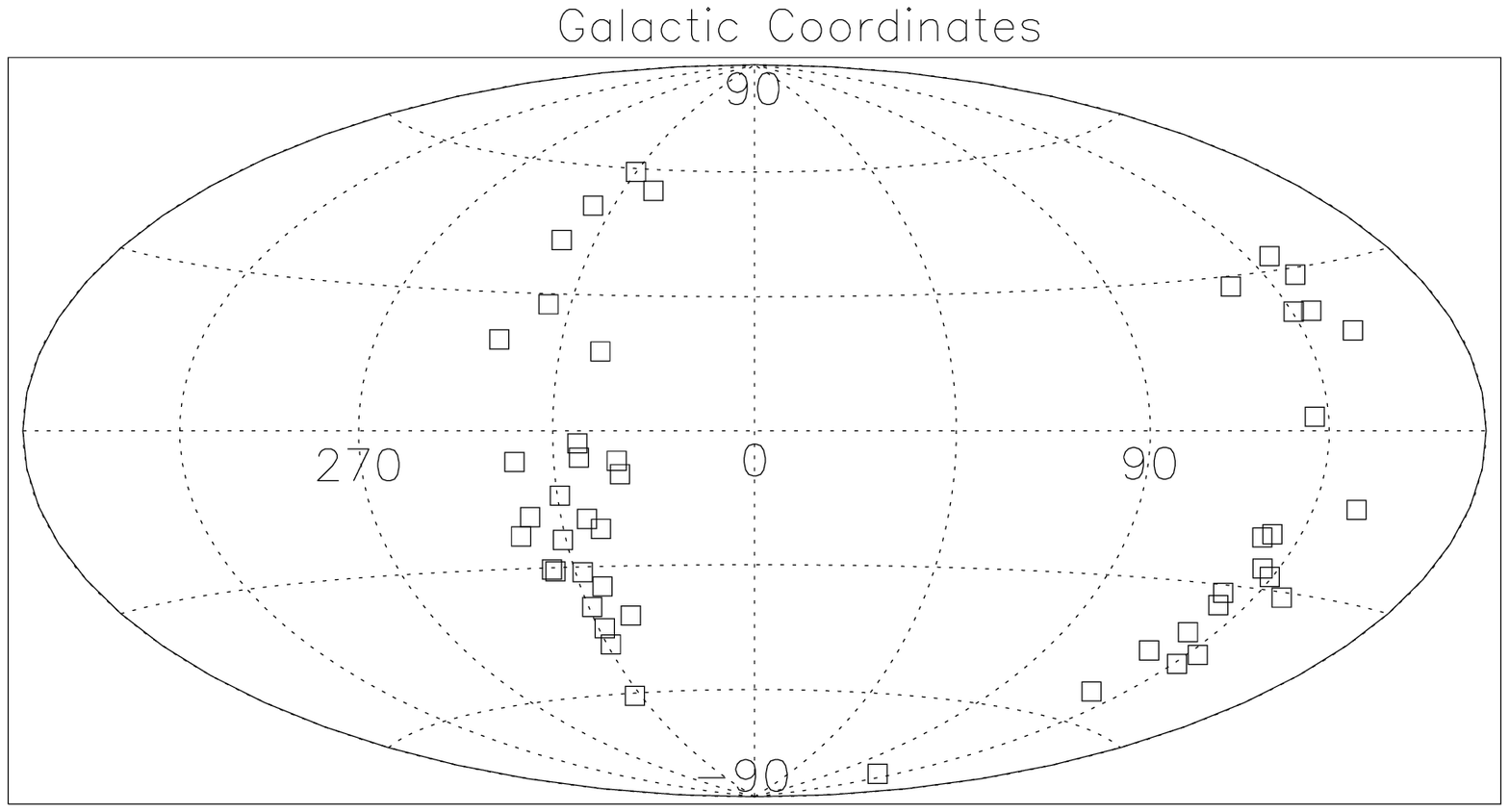}}
\epsfxsize=8.cm \centerline{\epsfbox{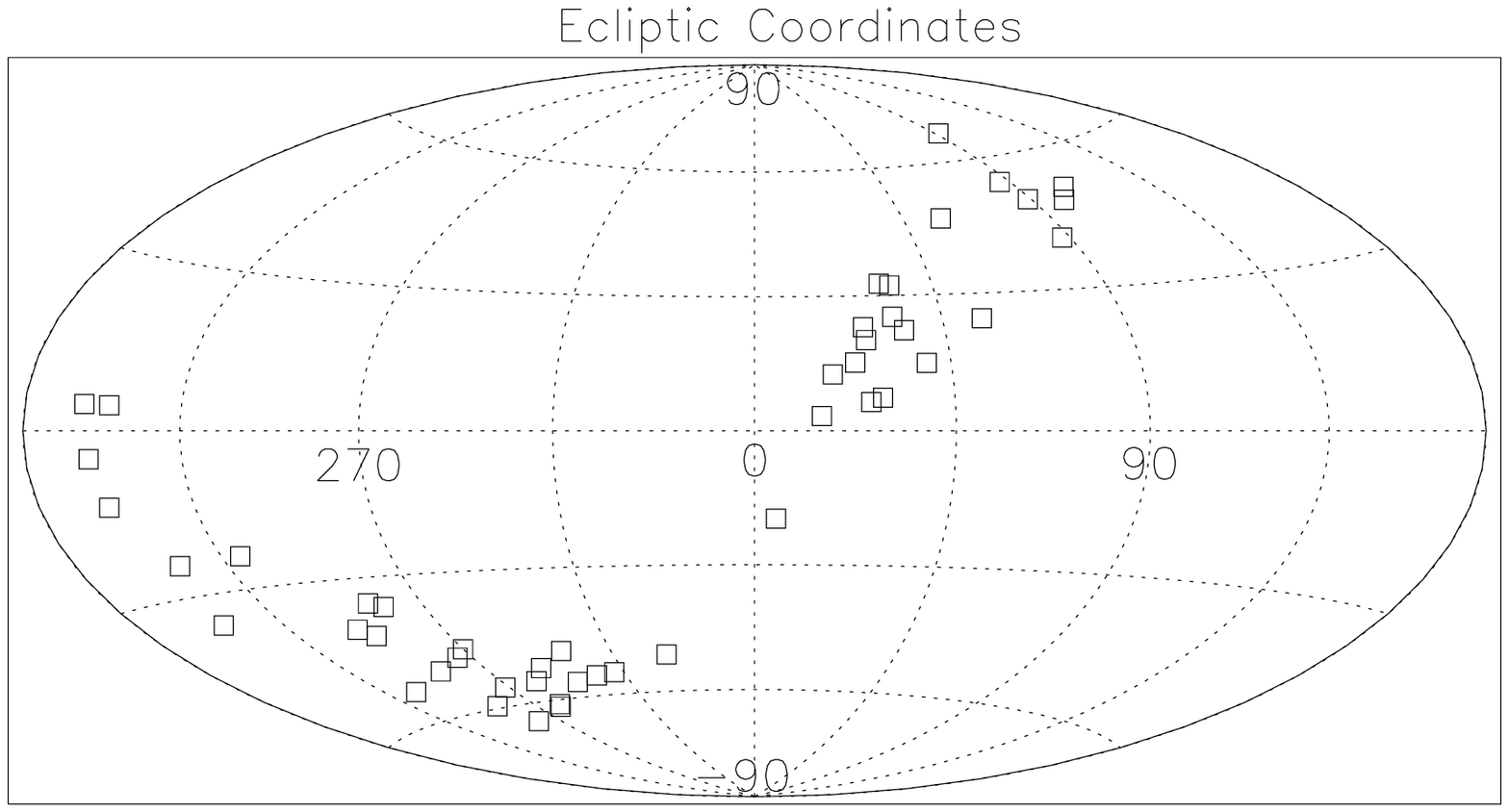}}
\caption{The plots show all 48 1A and 1B comets satisfying the
criteria $120^\circ \le \ell \le 150^\circ$ or $300^\circ \le \ell \le
330^\circ$.  The upper (lower) panel is a Hammer-Aitoff projection in
galactic (ecliptic) coordinates.}
\label{fig:matese}
\end{figure}

\subsection{The Planet of Matese, Whitman \& Whitmire}

Matese et al. (1999) argue that evidence for a massive ($\sim 3 \MJ$)
body moving in the Oort cloud can be seen in the aphelion positions of
1A and 1B comets, when plotted in galactic latitude and
longitude. They claim that there is an excess of objects in a great
circle band defined by $120^\circ \le \ell \le 150^\circ$ or
$300^\circ \le \ell \le 330^\circ$. Figure~\ref{fig:matese} shows the
great circle alignment in galactic and ecliptic coordinates.  The
stream is particularly obvious amongst dynamically new comets ($10^6/a
< 100$), i.e., comets that are passing through the Solar system for
the first time.  Matese et al. identify 30 out of the total of 82
dynamically new comets with orbits designated 1A or 1B in the 1996
version of Marsden \& Williams, which lie within the great circle
band. Of these, four have orbits that are apparently hyperbolic prior
to entering the inner solar system. (There are seven which also appear
in Murray's original list of objects).  The upper part of
table~\ref{table:matese} lists all comets originally identified by
Matese et al. (1999). The lower part lists all comets in Marsden \&
Williams (1999) satisfying the cut in galactic longitude alone.  Three
of these (C/1997 BA6, C/1999 F2 and C/1996 E1) are also dynamically
new.  The comets in the stream show good coverage of both the northern
and southern ecliptic hemispheres, which mitigates concerns about the
north-south bias. However, Matese et al.'s comet alignment is close to
normal to the solar apex, which is curious but perhaps coincidental.

\begin{figure}
\epsfxsize=8.cm \centerline{\epsfbox{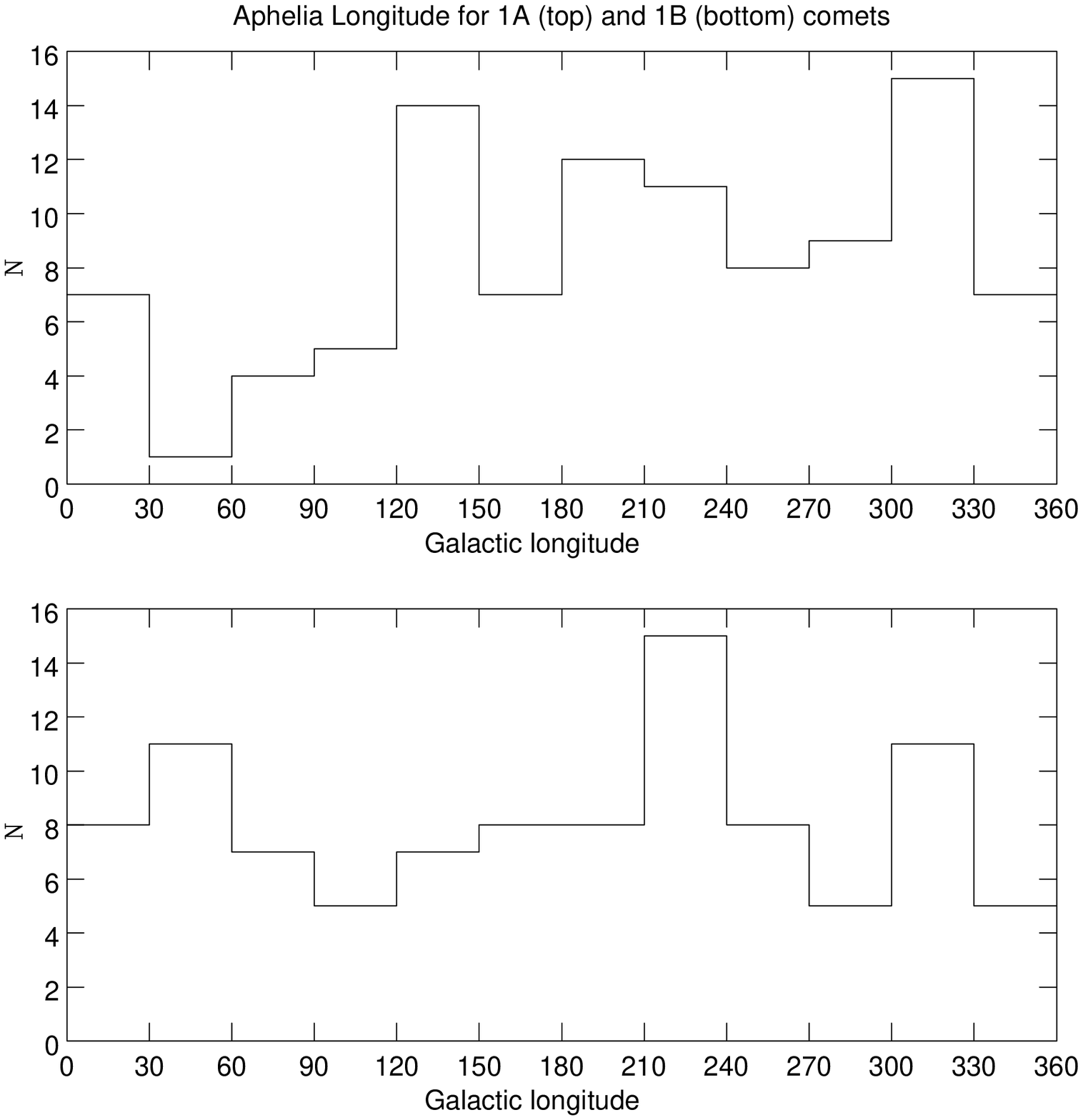}}
\epsfxsize=8.cm \centerline{\epsfbox{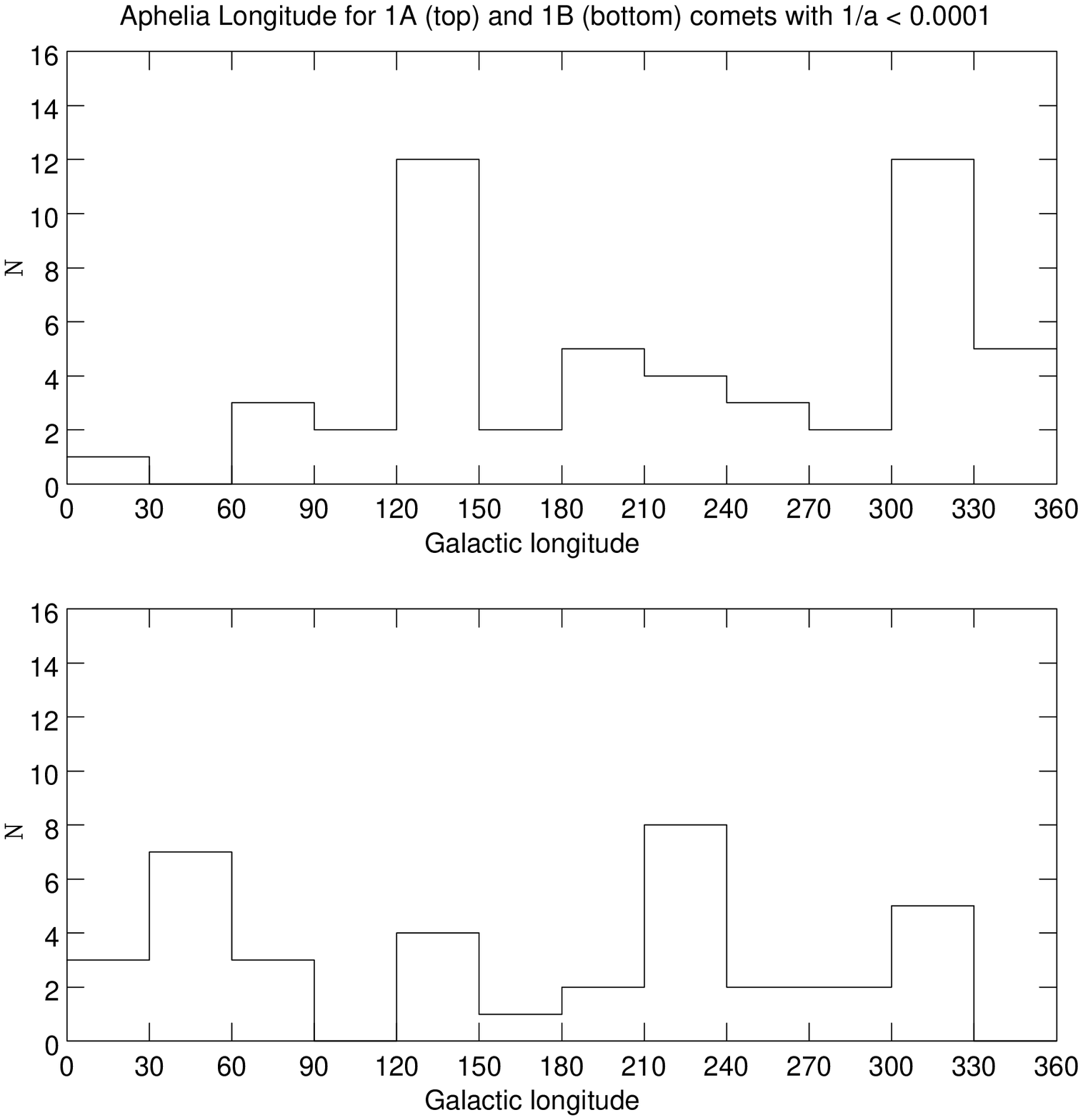}}
\caption{The histograms show the distribution in galactic longitude of
all 198 1A and 1B comets (upper two panels) and solely those 88
fulfilling the energy cut $1/a \lta 0.0001$ (lower two panels). Notice
that the great circle excess ($120^\circ \le \ell \le 150^\circ$ and
$300^\circ \le \ell \le 330^\circ$) is scarcely visible when all
comets are considered, but it is very obvious when only dynamically
new comets with the highest quality (1A) orbits are considered. (We
have omitted the cometary fragments C/1996 J1-A and J1-B from the
sample of dynamically new comets.)}
\label{fig:histmat}
\end{figure}
\begin{figure}
\epsfxsize=8.4cm \centerline{\epsfbox{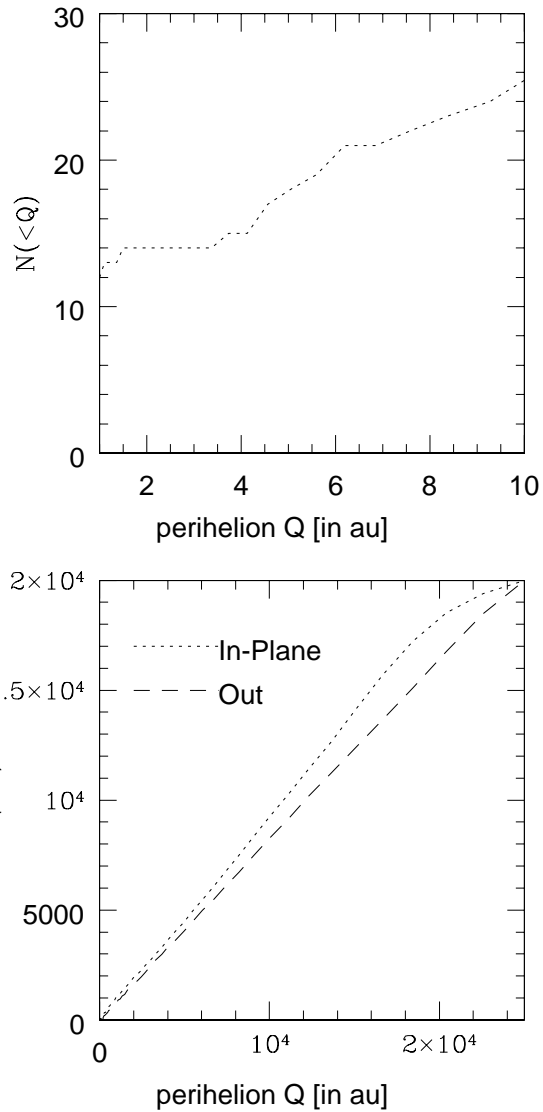}}
\caption{The cumulative number of comets reaching a perihelion
distance of at least $Q$ during the course of the 30 Myr
simulation. The upper panel shows a detail of the inner part of the
Solar System, into which only the in-plane comets penetrate. The lower
panel shows all the simulation data on the in-plane and out-of-plane
comets.}
\label{fig:simulation}
\end{figure}

Figure~\ref{fig:histmat} shows histograms of the aphelia of the 1A and
1B comets as a function of galactic longitude. The upper two panels
show all 1A and 1B comets with no cut placed on the value of
energy. The lower two panels show the equivalent histograms for the
dynamically new comets ($10^6/a \lta 100$). Matese et al.'s (1999)
Figure~2 is a combination of the lower two plots. Without a cut-off in
energy, the excess is present but not prominent. However, in the
sample of new comets, and especially those whose orbits are best
known, the excess is very obvious. Considering only the dynamically
new objects, the stream contains 24 1A comets out of 51 and 9 1B new
comets out of 37 and covers a sixth of the sky. In order to determine
the significance, Monte Carlo simulations are run to assess how easily
such deviations may be attributable to chance. Of course, there is no
a priori basis for assuming that the great circle arc passes through
the galactic poles. Therefore, it is equally impressive if the
overpopulated great circle is at any orientation.  This suggests that
we ask for the frequency with which 33 or more out of 88 points occur
in any band covering a sixth of the sky. Using Monte Carlo
simulations, this occurs with a probability of $\sim 1.4 \times
10^{-2}$. Suppose however the sample is contaminated by a weak comet
shower (like the Biermann shower) and this is responsible for 6 of the
objects in the great circle alignment, which are therefore not
statistically independent.  In this case, we only require $28$ comets
or more out of 83 in the stream, which occurs with a probability of
$\sim 8.2 \times 10^{-2}$. However, to reproduce the data on the 1A
comets alone, we require 24 or more comets in the stream out of a
sample of 51. This arises by chance only $\sim 1.5 \times 10^{-3}$
times. Hence, it seems that the pattern in the data discovered by
Matese et al. (1999) is quite unusual.  Perhaps the most remarkable
thing is that the signature is strongest in the 1A comets.  The more
careful we are to ensure that the comets are first time entrants, then
the greater the statistical significance despite the decreasing
numbers.

We use simulations to study whether a massive object in the orbit
proposed by Matese et al. could give such a signal. Using the Mercury
program (Chambers 1999), Planet X with a mass of $3\MJ$ is placed on a
circular orbit around the Sun at a radius of 25000 au.  Massless
particles are then distributed in this two-body system.  Each particle
starts with a semi-major axis of 25000 au, the same as the planet. The
orbits of the objects are chosen to give one hundred clones in
eccentricity, varying in steps of 0.01 from 0 to 0.99. This seems
reasonable as objects in the Oort Cloud are able to attain high
eccentricities both because they were originally ejected and because
of the flexing of the Galactic tide.  The inclinations of the clones
are $0^\circ, 90^\circ, 180^\circ$ and $270^\circ$ with respect to the
planet's orbital plane. In other words, the clones are restricted to
lie either in the same plane as, or perpendicular to, the planet's
orbit.  Finally, the objects are scattered in longitude of pericentre
so that there are 100 clones spread evenly around the full range of
$360^\circ$. This gives a total of 40000 massless particles in the
simulation.  The timestep chosen for the simulation is $10\%$ of the
orbital period of Planet X, and the simulation is run for 30 Myr.

Figure~\ref{fig:simulation} shows the cumulative number of clones
reaching a perihelion distance less than a certain value during the
duration of the simulation. Let us regard objects attaining a
perihelion of less than 5 au as satisfying the requirements for
discovery as long period comets. During the course of the simulation,
$19$ clones -- all initially in the same plane as the planet's orbit
-- managed to reduce their perihelion to less than 5 au. However, no
object starting out of plane ever came close enough to the Earth to be
discovered as a comet. So, the rough probability that an in-plane
clone becomes visible over the course of 30 Myr is $\sim 0.1 \%$.
Given the numbers in Matese et al.'s comet stream, a new member is
discovered on average once every 10 years. In other words, $3 \times
10^6$ comets in the stream must be generated over 30 Myr. This
suggests that the parent population of comets on perturbable orbits is
$\sim 3 \times 10^{9}$ objects. This is quite reasonable, as it is
only $\sim 0.3 \%$ of the total number of comets ($\sim 10^{12}$) in
the whole of the Oort Cloud.  Although our simulation is crude, it
does unambiguously show two things. First, the planet on its own can
perturb comets in its orbital plane so that they become visible on
timescales of tens of millions of years.  Second, the parent
population required to maintain the observed flux of comets in the
stream is not wildly implausible. So, Matese et al.'s planet warrants
further and serious consideration.

\section{Conclusions}

The catalogue of long period comets (Marsden \& Williams 1999) is used
to investigate the independent claims of Murray (1999) and Matese,
Whitman \& Whitmire (1999) of evidence for a massive undiscovered body
(Planet X) in the Oort Cloud of comets. This leads to an examination
of the biases present in the catalogue, both returning to work by
earlier authors and also studying several other biases which have
apparently not been discussed previously.

The biases considered are as follows. The first is the north-south
bias.  Especially in the past, the excess of observers in the northern
hemisphere over the southern hemisphere has led to a noticeable excess
in those comets which could be observed from that hemisphere. Hence,
long period comets with northern perihelia (and hence southern
aphelia) predominate in catalogues.  A plot of the number of comets
discovered per decade from both hemispheres in the years preceding
1956 shows dips at the times of a number of cataclysms (such as the
First World War). This is a indication that the long period comet
catalogue is telling us about more than science. Nowadays, most comets
are discovered by automated searches. However, both LINEAR and NEAT
are based in the northern hemisphere and so this selection effect
still persists.  Second, there are a number of diurnal and seasonal
biases. These introduce selection effects into the catalogue according
to time of day or time of year. The time of day has an effect on
whereabouts in the sky an object can be discovered. Obviously, if the
comet is too close to the sun, it will not be seen. In addition, if
the comet is near opposition, it is likely to be further from the sun,
and hence fainter and harder to observe. This leads to minima in the
observed distribution of comet discoveries both toward the sun and,
less so, towards the position of opposition on the sky.  During the
months around midsummer, the amount of dark time available to an
observer is less than that around midwinter. If other things were
equal, this would in turn mean that a greater number of comets are 
discovered during the local winter than summer. However, this effect
is somewhat lessened by the fact that the weather tends to be far less
clement in the winter, and hence the amount of time to observe is
reduced.  Both these effects are clearly present in the comet
catalogue. They create changes in the distribution that act to
accentuate or to lessen the effects of other biases, such as
directional biases.  Comets are, in principle, harder to discover in
areas of the sky with a dense stellar or nebular background than in
sparser areas. However, this may well be countered by the fact that
such dense areas are observed more often, which in turn boosts the
likelihood of the object's discovery (e.g., comet Hale-Bopp). A number
of orbital biases necessarily afflict the catalogue.  Comets with
perihelia at very small distances spend much less time in the inner
solar system and hence may be more likely to be missed, unless they
are bright enough to observe when close to the Sun. Also, comets with
large perihelion distance, beyond that of the Earth, are most likely
to be discovered near to opposition, when the Earth-comet distance is
the smallest. A further type of bias is that imposed by the observers
themselves -- for example, many observers concentrate their attention
close to the ecliptic, and hence objects moving on high inclination
orbits may be under-represented in the comet catalogue.  In order to
understand better some of the selection effects, a mathematical model
has been developed which allows some of these effects to be
illustrated.

Murray (1999) argues that there is an accumulation of cometary aphelia
between 30000 and 50000 au and this is a piece of evidence for Planet
X. However, this can also be explained by the competing effects of
greater spatial volume, yet increasing depletion from Galactic tides
on moving outwards in the Oort Cloud, inevitably giving rise to a
maximum. Murray also identifies a possible great circle alignment in
the aphelia directions of comets and a saw-tooth trend of orbital
period against ecliptic longitude.  However, the most recent data does
not maintain these patterns.  There is also evidence that the
subsample in the great circle stream is affected by seasonal and
north-south biases.

The stream of Matese et al. (1999) appears to stand up reasonably well
to critical examination. Considering only first time entrants to the
Solar system, the stream contains 33 comets out of 88 and covers a
sixth of the sky. This great circle excess of comets cannot be
attributed to any known selection effects.  Such an excess also seems
unlikely on probabilistic grounds.  The frequency with which 33 or
more out of 88 points occur in any band covering a sixth of the sky is
$\sim 1.4 \times 10^{-2}$.  To reproduce the data on the 1A comets
(i.e., those with the most accurate orbits) is a much stiffer
task. There are 24 1A comets out of 51 in the great circle stream.
This occurs with a probability of only $\sim 1.5 \times 10^{-3}$ by
chance. Hence, it seems clear that the pattern in the data discovered
by Matese et al.  (1999) is unusual. There are obvious concerns
regarding the small dataset involved. But, it is reassuring that the
trend identified by Matese et al. has been maintained with the
addition of newer data. It is also reassuring that the more careful we
are to ensure that the comets are first time entrants, the better is
the statistical significance of the stream.  Matese et al. advocate
the existence of an additional planet of mass $3\MJ$ possibly on a
circular orbit with a radius of 25000 au. Numerical integrations
demonstrate that such a planet by itself can perturb objects in its
orbital plane and reduce their perihelia to sufficiently small values
($Q < 5$ au) so that they could be discovered as long period comets
from the Earth. To maintain the observed flux of comets in the stream
requires a parent population of $\sim 3 \times 10^{9}$ on perturbable
orbits close to the planet's orbital plane. This seems a plausible
number, as it is only $\sim 0.3 \%$ of the total number of comets in
the Oort Cloud.  We conclude that Matese et al.'s planet is a
possible, perhaps even likely, explanation of the unusual pattern in
the data. The only alternative is to ascribe the pattern to an unlucky
artefact of the complex observational biases.

Clearly, there is a need for a sample of long period comets that is
free from unknown or hard-to-model selection effects. The European
Space Agency satellite GAIA (http://astro.estec.esa.nl/gaia/),
targetted for launch in 2010, will repeatedly survey the whole sky
complete to 20th magnitude during its 5 year mission lifetime. The
scanning law is complicated but readily modelled with a computer. A
slow-moving solar system object is typically seen a few hundred times
by GAIA during the mission lifetime, provided it is brighter than 20th
magnitude. We can estimate roughly the number of long period comets
that will be discovered by GAIA by scaling the experience of LINEAR,
which has a very similar limiting magnitude. LINEAR has found 59
comets since 1998, averaging $\sim 20$ per year.  We estimate that
over the course of 1 year, LINEAR covers $\lta 1/10$ of the sky. GAIA
covers the whole sky repeatedly over the year, so may be expected to
discover $\sim 20 \times 10 \times 5 = 1000$ long period comets during
the mission lifetime.  Such a dataset would allow the existence of the
stream of Matese et al. (1999) to be confirmed or denied. If such a
stream does indeed exist in a comet sample with easy-to-model biases,
then inferences may be drawn on the perturber, which may eventually
lead to its discovery.

\section*{Acknowledgments}
JH acknowledges financial support from PPARC, while NWE thanks
the Royal Society. Helpful conversations with Mark Bailey were
much appreciated. Both John Matese and John Murray gave us useful
comments on the draft manuscript.

{}

\label{lastpage}

\begin{thebibliography}{99}

\bibitem[Anderson \& Standish 1986]{as} Anderson J.D., Standish
E.M. 1986, In: ``The Galaxy and the Solar System'',
eds. R. Smoluchowski, J.N. Bahcall, M.S. Matthews, (University Of Arizona
Press: Tucson), p. 286

\bibitem[Biermann, Huebbner \& Lust]{BHL} Biermann L., Huebbner W.F.,
L\"ust R.H. 1983, Proc. Nati. Acad. Sci., 80, 5151

\bibitem[Chambers(1999)]{1999MNRAS.304..793C} Chambers, J.E. 
1999, MNRAS, 304, 793

\bibitem[Danby 1992]{danby} Danby J.M.A., 1992, Celestial Mechanics,
(Willman-Bell: Richmond)

\bibitem[Davis, Hut, \& Muller(1984)]{1984Natur.308..715D} Davis M.,
Hut P., Muller, R.A. 1984, Nature, 308, 715

\bibitem[EFS 2002]{efs} Evans N.W., Ferrer F., Sarkar S., 2002,
Astropart. Phys., 17, 319 

\bibitem[Everhart 1967]{edgar} Everhart E. 1967, AJ, 72, 716

\bibitem[Forbes (1880)]{forbes} Forbes G. 1880, The Observatory,
3, 439

\bibitem[Guliev (1992)]{guliev} Guliev A.S. 1992, Soviet Ast.
Lett. 18, 75

\bibitem[Holetschek 1891]{holetschek} Holetschek J. 1891, Astron. 
Nachr. 126, 75

\bibitem[Hogg et al]{} Hogg D.W., Quinlan G.D., Tremaine S.D., 1991,
AJ, 101, 2274

\bibitem[Hughes]{} Hughes D.W. 1983, MNRAS, 204, 23

\bibitem[Hughes]{} Hughes D.W. 2001, MNRAS, 326, 515

\bibitem[Jaschek \& Valbousquet 1994]{1994A&A...291..448J} Jaschek C.,
Valbousquet A. 1994, AA, 291, 448

\bibitem[Kenyon \& Luu 1999]{kenyonluu} Kenyon S.J., Luu J.X. 1999,
AJ, 118, 1101

\bibitem[Kresak 1982]{kresak} Kres\'ak L. 1982, In ``Comets'',
ed L.L. Wilkenning (University of Arizona Press: Tucson), p. 56

\bibitem[Kristensen 1998]{kris} Kristensen L.K., 1998, Astron. Nachr.,
319, 193

\bibitem[Kritzinger 1954]{kritziner} Kritzinger H.H. 1957,
Nachrichtenblatt der Astronomische Zentralstelle, 11, 4

\bibitem[42]{l75} Linsley J., 1975, Phys. Rev. Lett., 34, 1530.

\bibitem[L\"ust 1984]{rhea} L\"ust R.H., 1984, AA, 141, 94 

\bibitem[Lyttleton 1953]{ray} Lyttleton R.A., 1953, The Comets and
Their Origin (Cambridge University Press: Cambridge)

\bibitem[Marsden 1989]{1989AJ.....98.2306M} Marsden B.G. 1989, AJ,
98, 2306

\bibitem[Marsden \& Williams 1999)]{mw} Marsden B.G., Williams
G.V., 1999, Catalogue of Cometary Orbits, Smithsonian Astrophysical
Observatory, Cambridge, Mass.

\bibitem[Matese \& Whitmire(1986)]{1986Icar...65...37M} Matese J.J., 
Whitmire D.P. 1986, Icarus, 65, 37 

\bibitem[Matese, Whitman, \& Whitmire 1998]{mww} Matese
J.J., Whitman P.G., Whitmire D.P. 1998, Cel. Mech., 69, 77

\bibitem[Matese, Whitman, \& Whitmire 1999]{1999Icar..141..354M} Matese
J.J., Whitman P.G., Whitmire D.P. 1999, Icarus, 141, 354 

\bibitem[McKinley]{mckinley} McKinley D.W.R., 1961, Meteor Science and
Engineering, (McGraw-Hill: New York)

\bibitem[Murray 1999]{1999MNRAS.309...31M} Murray J.B. 1999, MNRAS, 
309, 31 

\bibitem[Neslusan 1996]{neslusan} Neslu{\v s}an L., 1996, AA, 306, 981, 

\bibitem[Opik 1971]{1971IrAJ...10...35O} Opik E.J. 1971, Irish 
Astron. J., 10, 35 

\bibitem[Stern 1991]{1991Icar...90..271S} Stern S.A. 1991, 
Icarus, 90, 271 

\bibitem[Tremaine 1986]{st} Tremaine S.D. 1986, In: ``The Galaxy and
the Solar System'', eds. R. Smoluchowski, J.N. Bahcall, M.S. Matthews,
(University Of Arizona Press: Tucson), p. 409

\bibitem[Vandervoort \& Sather 1993]{vandervoort} Vandervoort 
P.O., Sather E.A. 1993, Icarus, 105, 26 

\bibitem[Vsekhsvyatskii 1964]{v} Vsekhsvyatskii 1964, The Physical
Characteristics of Comets (Israeli Program for Scientific
Translations: Jerusalem)

\bibitem[Whitmire \& Jackson 1984]{1984Natur.308..713W} Whitmire D.P.,
Jackson A.A. 1984, Nature, 308, 713 

\bibitem[Whitmire \& Matese]{wm} Whitmire D.P., Matese J.J. 1985,
Nature, 313, 36

\bibitem[Yabushita]{sy} Yabushita S., 1979, MNRAS, 189, 45

\end{thebibliography}
\end{document}